\documentclass{aastex631}
\usepackage{graphicx} 
\usepackage{amsmath}

\newcommand\nasagoddard{
NASA Goddard Space Flight Center, Greenbelt, Maryland, USA
}

\newcommand\stewardobservatory{
Steward Observatory and Department of Astronomy, The University of Arizona, Tucson, AZ 85721, USA}

\begin{document}

\title{Bioverse: Assessing the Ability of Direct Imaging Surveys to Empirically Constrain the Habitable Zone via Trends in Albedo}

\author[0000-0003-3989-5545]{Noah W. Tuchow}
\affil{\nasagoddard}
\affil{\stewardobservatory}
\email{nwtuchow@arizona.edu}

\author{Christopher C. Stark} 
\affil{\nasagoddard}

\author[0000-0003-3714-5855]{D\'aniel Apai} 
\affil{\stewardobservatory}
\affil{Lunar and Planetary Laboratory, University of Arizona, 1629 E. University Boulevard, Tucson, AZ 85721, USA}
\affil{Alien Earths Team, NASA ICAR/NExSS, USA}

\author[0000-0001-8355-2107]{Martin Schlecker}
\affil{\stewardobservatory}
\affil{European Southern Observatory, Karl-Schwarzschild-Straße 2, 85748 Garching bei München, Germany}

\author[0000-0003-3702-0382]{Kevin K.\ Hardegree-Ullman}
\affil{\stewardobservatory}
\affil{Caltech/IPAC-NASA Exoplanet Science Institute, 1200 E.\ California Blvd., MC 100-22, Pasadena, CA 91125, USA}

\section*{Abstract}
Will future direct imaging missions such as NASA's upcoming Habitable Worlds Observatory (HWO) be able to understand Earth-sized planets as a population? In this study, we simulate the ability of space-based coronagraphy missions to uncover trends in planetary albedo as a function of instellation, and potentially constrain the boundaries of the habitable zone. 
We adapt the \texttt{Bioverse} statistical comparative planetology framework to simulate the scientific output of possible designs for HWO. With this tool, we generate a synthetic planetary population with injected population-level trends in albedo and simulate the observability of planets. We then determine the statistical power to which these trends can be recovered as a function of the strength of the injected trend and the sample size of Earth-sized planets in the habitable zone (exoEarths). The strongest trends in albedo require a sample size of roughly 25-30 exoEarths to recover with high confidence. However, for weaker albedo trends, the required number of planets increases rapidly.
If a mission is designed to meet the Decadal Survey's requirement of 25 exoEarths, it would be able to recover very strong trends in albedo associated with the habitable zone, but {would} struggle to confidently detect weaker trends. We explore multiple strategies to increase one's ability to recover weak trends, such as reducing the uncertainties in observables, incorporating additional observables such as planet colors, and obtaining direct constraints on planetary albedo from full spectral retrievals.  

\section{Introduction}

The search for habitable exoplanets is one of the main focuses of modern astronomy and astrophysics. We seek to {understand the nature} of potentially Earth-like planets to determine where habitable conditions emerge and where one might expect to find signs of life. A great deal of effort is being devoted to the search for Earth-sized planets in the habitable zone (hereafter exoEarth candidates or EECs). In the near future, astronomers aim to attain the sensitivity necessary to directly image rocky habitable zone exoplanets in reflected light. This goal is put forward in the Astro2020 Decadal Survey, which recommends a future large flagship mission with the ultimate goal of detecting and characterizing approximately 25 EECs via space-based high-contrast imaging \citep{DecadalSurvey}. 
NASA's upcoming Habitable Worlds Observatory (HWO) aims to meet the science goals set by the Decadal Survey. Slated to launch in the 2040s, HWO will be designed to recover the atmospheric compositions of individual exoplanets {with the aim of inferring} whether they are habitable or host biosignatures. 

However, it is not clear whether the recommended sample size of 25 EECs is sufficient to understand Earth-sized planets \textit{as a population}. Ultimately we seek to understand the physical processes that govern the habitability, formation, and evolution of rocky exoplanets. Close study of individual planet candidates can retrieve their atmospheric compositions and fundamental properties, but to better test our theories about the nature of exoplanets one needs to identify statistical trends in the planetary population \citep{Apai2017,Bean2017,Apai2019,Checlair2019}. This will enable us to obtain information that wouldn't be available from the study of individual planets alone.  

We seek to understand whether the Decadal Survey's goal of approximately 25 EECs will be sufficient to identify underlying trends in the planetary population. If not, what sample sizes are required? 
In this study, we develop a methodology to determine the sample size required to test a given hypothesis about the planet population. With this approach, we can inform which science cases are feasible to investigate via space-based direct imaging. We consider the specific example of HWO in our analysis, though we note that our results should be generalizable to any direct imaging survey with the sensitivity necessary to detect exoEarths.
As HWO is currently in the early phases of mission development, several possible designs are being considered. Trade studies are ongoing to assess their science performance, feasibility, and cost. Among the metrics used to evaluate the science output of potential designs, the yield of EECs is among the most prominent. This research will calculate the sample sizes required to recover population-level trends, which can then be used as a metric for engineering trade studies. 

\subsection{Albedo trends}
\label{hz_albedo_trends}
The primary science case we address in this study is whether a direct imaging mission such as HWO would be able to identify population-level trends in planetary albedo as a function of instellation (stellar incident flux). 
Uncovering the distribution of albedos for rocky planets may enable us to test our theories about planetary habitability. In particular the classical picture of the habitable zone (HZ) makes qualitative predictions about the albedos of Earth-like planets.  The concept of the HZ defines a range of instellations around a star where a planet with an Earth-like atmosphere could maintain liquid water on its surface \citep{Kasting1993hz,Kopparapu2013}. The boundaries of the HZ quantify the instellations at which the stabilizing climate feedbacks, such as the $\mathrm{CO}_2$ cycle, break down. Assuming that rocky planets in the habitable zone are subject to similar climate regulating processes to the Earth, one would expect that Earth-like planets in the HZ should have lower albedos than those outside of it. 
This may seem somewhat counterintuitive, but it makes sense when considering planets outside the HZ where the regulating climate feedback cycles break down. 
Planets interior to the inner edge of the HZ will have thick runaway greenhouse atmospheres with highly reflective cloud decks, similar to Venus in our solar system. Planets beyond the outer edge will reach the maximum greenhouse limit where adding additional $\mathrm{CO}_2$ to the atmosphere will not keep the planet above the freezing point of water \citep{Kopparapu2018review}. Such planets may have thick atmospheres and would experience global glaciation and the freeze out of water and eventually $\mathrm{CO}_2$ from their atmospheres, further trapping them in a glaciated state with high albedo (perhaps similar to icy bodies in the outer solar system). In comparison to planets outside the HZ, the HZ is a region of lower planetary albedo where Earth-like planets have less thick and cloudy atmospheres and absorptive surface oceans. 

We have a heuristic understanding about how planetary albedos will vary inside and outside the HZ, but to the best of our knowledge, models have yet to predict the distribution of albedos as a function of instellation.
This would require an understanding of the diversity of compositions of planets in the HZ and their occurrence rates which are unlikely to be attainable prior to HWO. If done today, such models would need to rely on simplifying assumptions. Nonetheless, even if the current state of theoretical research isn't able to produce exact numerical estimates, {future missions may be able to infer the albedo distribution of planets and thus place empirical constraints on the HZ boundaries.} 

\subsection{Inferring planetary albedo}

{ When imaging planets in reflected light, the albedo plays a critical role in determining their brightness and thus their detectability. Trends in albedo will manifest themselves as trends in planet-star contrast, which can be directly measured from planet detections alone. Contrasts will be much easier to measure than other planetary properties, such as molecular concentrations, which will require full spectral characterization of the planet, requiring much more exposure time. Retrievals of the atmospheric properties of rocky exoplanets will only be available for a small sample of well studied exoplanets. As the spectra of these planets will likely be low signal-to-noise, retrievals may have a difficult time constraining their atmospheric compositions. Identifying trends in albedo potentially bypasses these problems, at it relies on measurements from planet detections alone, which will be a much larger sample than fully spectrally characterized exoplanets.}

In reflected light, the planet-star contrast will be directly proportional to the planetary geometric albedo. Contrast is given by 
\begin{equation}
    \label{eq1}
    C= \frac{F_{\mathrm{p}}}{F_{\mathrm{s}}}=A_\mathrm{g}\left(\frac{R_\mathrm{p}}{a}\right)^2\Phi(\alpha)
\end{equation}
where {$F_\mathrm{p}$} and {$F_\mathrm{s}$} are the fluxes from the planet and star respectively,  $A_\mathrm{g}$ is the planetary geometric albedo, {$R_\mathrm{p}$} is the planet radius, $a$ is the semi-major axis, and $\Phi(\alpha)$ is the phase function for phase angle $\alpha$ {\citep{Traub2010,madhusudhan2012}}. The albedo distribution for rocky planets is fundamentally unconstrained, but it will strongly affect the observability of planets via direct imaging. This makes it difficult to calculate the yield of planets HWO will detect without making assumptions about the albedo distribution. Conversely, because of the role albedo plays in the detectability of planets, it may be possible to uncover the underlying distribution of planetary albedos by analyzing the distribution of observed contrasts.

Albedo will not be directly observable via high contrast imaging but can be inferred from measured contrasts. The main observables from a direct imaging detection are the planet-star contrast and angular separation. With multi-epoch observations one can measure a planet's semi-major axis, and given knowledge of the host star luminosity, $L$, one can calculate instellation as $S_\mathrm{eff}=\frac{S}{S_\oplus}=\frac{ L/L_\odot}{\left(a/1 \mathrm{AU}\right)^2}$.
As we are interested in trends in albedo or equivalently contrast as a function of instellation, we can rewrite the equation for contrast in terms of $S_\mathrm{eff}$:
\begin{equation}
    C = A_\mathrm{g} \left(\frac{R_\mathrm{p}}{1 \mathrm{AU}}\right)^2 S_\mathrm{eff} \left(\frac{L}{L_\odot}\right)^{-1}\Phi(\alpha)
\end{equation}
Multiple terms in this equation for contrast will be known for a direct imaging detection. Dividing out these terms we obtain a quantity that we refer to as $\beta$, which is the closest directly observable property to albedo:

\begin{equation}
    \beta = C \left(\frac{L}{L_\odot}\right)\frac{1}{S_\mathrm{eff}}= A_\mathrm{g} \left(\frac{R_\mathrm{p}}{1 \mathrm{AU}} \right)^2 \Phi(\alpha)
\end{equation}
$\beta$ can be thought of as a normalized contrast value, i.e. what the contrast would be if the planet were at 1 AU.
One can clearly see that this quantity represents a known degeneracy between planetary radius, albedo, and phase function {\citep{Cahoy2010}}. 
Breaking this degeneracy may be possible with precise atmospheric retrievals and multi-epoch observations, but for the purposes of this study we assume a scenario where one does not know the radius or phase of the planets that are detected.

A detection of an exoplanet via direct imaging without full spectral characterization would still yield measurements of planet-star contrasts and angular separations. We seek to understand whether it is possible to constrain the underlying albedo distribution using these observables, and whether the sample sizes required would be attainable with potential HWO architectures.
In this study, we generate a simulated population of exoplanets and model their observability with an HWO-like mission. We inject trends in planetary albedo to this population and forward model the distribution of planets in contrast (or $\beta$) and $S_\mathrm{eff}$. We vary the strength of this trend and the number of exoEarths in our sample and determine the statistical power to which this trend can be recovered.


\section{Methods}

To assess the ability of proposed direct imaging mission designs to detect population level trends in albedo we use the \texttt{Bioverse} statistical framework {\citep{Bixel2021,HardegreeUllman2023,Schlecker2024, HardegreeUllman2025,Schlecker2025}}. \texttt{Bioverse} is a statistical comparative planetology code that models the ability of future exoplanet surveys to uncover statistical trends in an observed planetary population. 
It consists of three main modules: Planet generation, survey simulation, and hypothesis testing. Each of these modules can be easily modified allowing users to specify the trend they are injecting, the observation method and instrument for the survey, and the hypothesis that they are testing. Individual steps in the planet generation process can be swapped out or added as necessary, allowing users a large degree of freedom over the universe of synthetic planets that are simulated. 


\texttt{Bioverse} has been used for past studies focusing on multiple detection methods and observatory designs ranging from ground-based ELTs, to space-based transit spectroscopy surveys, to proposed future direct imaging concepts such as LUVOIR-A \citep{Bixel2021,HardegreeUllman2023,Schlecker2024,HardegreeUllman2025}. In this study we shall expand the capabilities of \texttt{Bioverse} to simulate proposed HWO designs and improve the fidelity of its high contrast imaging simulation.



\subsection{Changes to Bioverse}

We modify the existing \texttt{Bioverse} code, making use of existing functionalities when possible and integrating new features. These changes are publicly available in our fork of the \texttt{Bioverse} GitHub repository\footnote{\url{https://github.com/nwtuchow/bioverse}}. Unless specified, our simulations for an exoEarth survey use the default settings for a direct imaging survey in \texttt{Bioverse}. 
For instance, we use the \citet{Kopparapu2014} formulation of the HZ and the \citet{Kopparapu2018} planet classification scheme.
To generate planets around host stars we used the ExoPAG Study Analysis Group 13 (SAG 13) occurrence rates modified to include the dependence of planet occurrence on spectral type \citep{Mulders2015a,Mulders2015b,Pascucci2019,Neil2020}. We use a value for the occurrence rate of EECs, $\eta_\oplus=0.24$, as given by SAG 13. This value has been used as a reference value of $\eta_\oplus$ for yield calculations for the HabEx and LUVOIR mission concepts, allowing for the comparison of results with different yield modeling codes \citep{stark2019,morgan2019}. More recent work by \citet{Bryson2021}, modeled the occurrence rates of EECs using the the same definition of an exoEarth as used by HabEx, LUVOIR, and this current study. Calculating occurrence rates in terms of instellation rather than orbital period, they found that $\eta_\oplus$ was bounded between 0.18 and 0.28 in the conservative HZ, which is broadly consistent with our adopted value of 0.24.  Estimates of $\eta_\oplus$ vary widely throughout the literature, but our results will be very insensitive to the value of $\eta_\oplus$ and will depend more strongly on the shape of the assumed planetary distribution function.
As we will address in Section \ref{survey_dists_section}, we are primarily concerned with accurately simulating the distribution of planets that would be detected, rather than predicting the absolute yield of EECs from any specific mission design. We provide constraints on the required sample sizes of EECs to recover albedo trends, while $\eta_\oplus$ will primarily have an effect on yield simulations to assess {whether} mission designs will be able to obtain that sample size.

Beyond the existing options for planet generation and survey simulation in \texttt{Bioverse}, we have made a series of changes to introduce new functionalities. We describe these changes below:

\subsubsection{Target lists and prioritization}
We have modified the stellar target list used for direct imaging simulations and have introduced a new prioritization scheme.  Earlier versions of \texttt{Bioverse} used a target list of the highest priority stellar targets for the 15m LUVOIR-A mission concept \citep{Bixel2019}. This target list was informed by yield calculations, identifying stars that would be selected for survey on the basis of the detectability of Earth-like planets in the HZ \citep{LUVOIR_final_report}. However, this target list was generated using assumptions about telescope and coronagraph designs specific to LUVOIR-A. We know from past yield studies that different telescope and coronagraph designs can select vastly different populations of stars for survey \citep{Stark2014}. Therefore, in order to simulate the results of other direct imaging mission designs, a different stellar sample will be required for each one. 

To ensure that the target list used in \texttt{Bioverse} does not contain assumptions that are specific to a given telescope design, we developed a methodology to allow \texttt{Bioverse} to generate a list of stars to be observed based on the specified mission architecture. We start by importing the HWO Preliminary Input Catalog (HPIC), containing a large sample of roughly 13,000 bright nearby stars that are potential HWO targets \citep{Tuchow2024HPIC}. This catalog contains all potential HWO targets spanning the range of possible mission designs.

To identify which stars will be observed by a given direct imaging mission, we need a target prioritization scheme. We developed a simple target prioritization scheme for direct imaging using \texttt{Bioverse}, selecting targets based on the exposure time necessary to characterize exoEarth atmospheres. In Section \ref{exposure_time_calc} we discuss how we develop an exposure time calculator that can be used for a wide variety of mission designs. We use this exposure time calculator to find the required exposure times to spectrally characterize an Earth analog in the HZ for each of the stars in the HPIC. Each of these hypothetical Earth analogs is assumed to be at Earth equivalent instellation distance around their star and is observed at quadrature phase. Exposure times are calculated for spectral characterization at 550 nm with SNR=10 and R=140.  We select target stars based on required exposure time, starting from the lowest values until the sums of exposure times exceeds a specified survey duration. For this study we assume a survey duration of 2 years {(40\% of a nominal 5 year mission)}, not including overheads and slew times, and {we} schedule observations of target stars until the total survey duration is exhausted.

With the results of our target prioritization scheme we replace the function to read the LUVOIR-A target list with a function to read the rows of the HPIC for the stars that we have selected.

\subsubsection{Phase calculation}
Previous \texttt{Bioverse} direct imaging simulations assumed that all planets were observed at quadrature phase. However, for this study we want to move beyond this simplifying assumption and simulate planets distributed over a more realistic range of planetary phases, taking an approach similar to that of \citet{HardegreeUllman2025}. We generate planets at random orientations around their stars with orbital elements generated using the existing \texttt{Bioverse} `assign\_orbital\_elements' function. For each planet we solve Kepler's equation, calculating the planet's phase angle and angular separation from its host star. Using Equation \ref{eq1} we calculate the planet-star contrast, assuming a Lambertian phase function. We acknowledge that the measured phase functions of solar system planets often deviate from ideal Lambertian spheres, but modeling the theoretical distribution of phase functions for a population of synthetic planets is beyond the scope of this study, and this rough assumption will suffice for our purposes. 

\subsubsection{Exposure time calculator}
\label{exposure_time_calc}

To aid in the target prioritization process {and improve} our detection criteria, we have added the ability for \texttt{Bioverse} to calculate the exposure times required to characterize exoplanet atmospheres. 
We have developed a simple exposure time calculator based on the procedure developed by \citet{Stark2014} (see their section 5.3, equations 7 - 15).  This exposure time calculator determines the time required to get a detection at a specified signal-to-noise as a function of wavelength, bandwidth, planet-star contrast, angular separation and host star magnitude. This exposure time calculator includes a simple model of a coronagraph with a step function for contrast between the coronagraph inner working angle (IWA) and outer working angle (OWA). Many telescope and coronagraph parameters can be varied such as the telescope diameter, coronagraph IWA, OWA, and minimum achievable contrast (a proxy for the noise floor). {Our exposure time calculator also takes into account the amount of exozodiacal light per star in units of the solar system's zodiacal light (zodi). For this study, we assume an average value of 3 zodi per star informed by the median amount of measured exozodiacal emission from the HOSTS survey \citep{Ertel2020}.}

We have updated the \texttt{Bioverse} survey module to use this exposure time calculation as a detection criterion to determine the yield of exoplanets detected by a survey. For every simulated exoplanet we calculate the exposure time needed for an SNR=7 detection.
If the time required to detect a planet is shorter than the amount of observation time allocated to its host star, we consider the planet to be detected.

\subsection{Injecting trends}

As discussed in Section \ref{hz_albedo_trends}, the classical picture of the HZ predicts that Earth-like planets in the HZ will have lower albedos than those outside of it. 
However, at our current level of understanding we lack quantitative predictions about what the distribution of planetary albedos will be inside and outside the HZ. We hope to be able to use direct imaging observations to uncover the unknown albedo distribution and learn about Earth-sized planets as a population. 
In this study, we seek to determine whether proposed HWO designs will be able to differentiate between multiple models for albedo as a function of $S_\mathrm{eff}$ and determine which of them best fits the data. 

For our purposes we shall use a toy model of albedo in the HZ, reflecting the qualitative predictions from models of the HZ and solar system observations. Our model of planetary geometric albedo takes the following form:
\begin{equation}
    A_\mathrm{g} = 
    \begin{cases}
        A_0 - \Delta A & \text{if in HZ,  }R_p<1.4 R_\oplus \\
        A_0 & \text{Otherwise}   
    \end{cases}       
\end{equation}
Here planets outside the HZ have an albedo of $A_0$, and rocky planets in the HZ have a lower albedo, with the difference in albedos inside and outside the HZ given by $\Delta A$. 
We use the \citet{Kopparapu2013} formulation of the HZ boundaries and adopt 1.4 $R_{\oplus}$ as the upper radius limit for planets considered to be exoEarth candidates as defined in previous studies such as \citet{Kopparapu2018}.

In our analysis, we shall first focus on injecting and recovering a strong trend in albedo where $A_0-\Delta A = 0.3$ and $A_0 = 0.7$.
This is similar to Earth albedo in the HZ and Venus albedo outside the HZ {\citep{mallama2017, Robinson2025}}. This relatively strong trend serves as a good test case to compare to a scenario with no injected albedo trend. Our ``null hypothesis'' will be a scenario where all planets have the same albedo value, i.e. $\Delta A =0$. Later, once we have established that it is possible to recover the strongest trend, we will reduce the strength of the injected trend parameterized by $\Delta A$ and see how that affects the required sample sizes needed to recover the trend.

\subsection{Measurement Noise}
\label{noise_section}

In order to simulate a more realistic output for a direct imaging survey, we incorporate estimates of the measurement uncertainties present in the dataset. We generate Gaussian white noise for measurements of the stellar luminosity $L$, planet semi-major axis $a$, and planet-star contrast $C$. We then propagate the uncertainties for these quantities to obtain the uncertainty in derived properties such as $S_\mathrm{eff}$ and $\beta$. 

{When generating Gaussian noise for observations, we use a characteristic value for the uncertainty in each quantity.} Here we describe our choices for these characteristic uncertainties.  For stellar luminosity, \citet{tayar2022} report a characteristic value for the minimum achievable precision for $L$ of $2.4\pm0.6\%$ due to uncertainties introduced by bolometric fluxes, photometric zero points, and atmospheric models among other factors. Of course most stellar luminosities will not be measured as precisely as this effective limit. To gain a sense of the characteristic $L$ uncertainty for HWO target stars we looked at the values listed in the NASA Exoplanet Exploration Program (ExEP) Mission Stars list of 164 best targets for HWO \citep{mamajek2023}. In the ExEP list we observe that the average uncertainty in $L$ is $2.94\%$, so for our purposes we shall use a luminosity uncertainty of $3\%$.  For the planet-star contrast, the uncertainty is motivated by the signal-to-noise required for detection and is given by 1/SNR. In our exposure time calculator, we use an SNR of 7 as a criterion for planet detection, consistent with the value used by \citet{stark2015}. This SNR gives us a $14\%$ uncertainty for contrast measurements. The uncertainty in semi-major axis is somewhat difficult to estimate and is based on the ability of direct imaging observations to constrain the planet's orbit by observing it at multiple epochs. Figure 3.1-2 in the HabEx Final Report gives the semi-major axis uncertainty as a function of the number of detections of a planet over multiple epochs \citep{gaudi2020}. For a circular orbit, the uncertainty in $a$ can be reduced to around $10\%$ after 3 detections, dropping down to $1\%$ after 4 detections. We note that our current simulation does not take into account multiple visits to a target and multi-epoch observations, so to gain a constraint for the value of uncertainty in $a$ we look toward the survey requirements. In order to perform accurate spectral retrievals for rocky planets in reflected light one requires a value of $S_\mathrm{eff}$ to be characterized to within $10\%$ uncertainty. This imposes a roughly $5\%$ uncertainty requirement for $a$, which we adopt as our characteristic value for semi-major axis uncertainty. As discussed earlier, such an uncertainty in $a$ could be achieved using between 3-4 detections well spaced over multiple epochs. 

\subsection{Survey Simulation}
\label{survey_sim}
\begin{figure}
    \centering
    \plotone{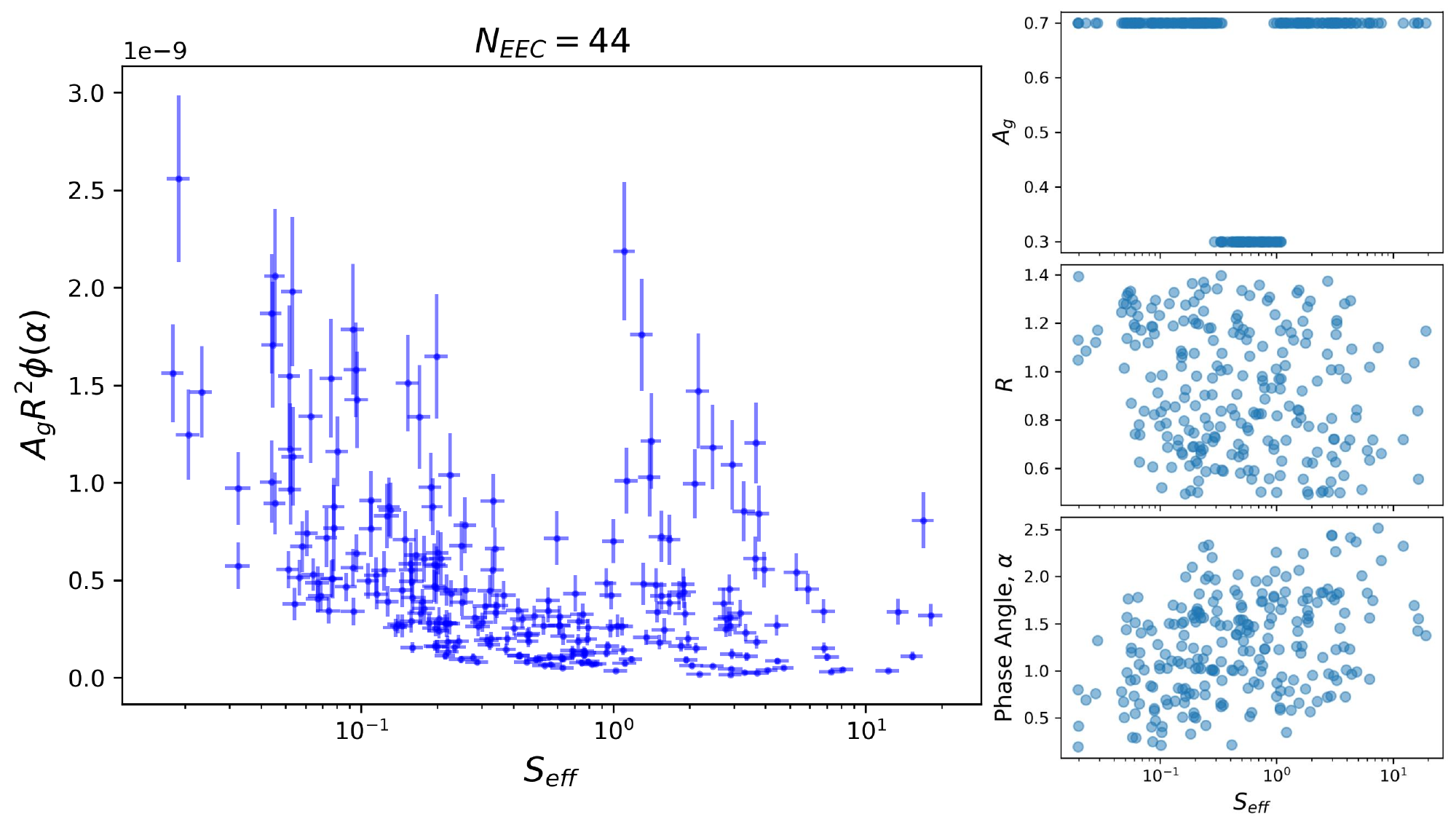}
    \caption{Example \texttt{Bioverse} simulation of a survey to characterize Earth-sized planets in the HZ. This simulation used an 8m diameter telescope observing at $\lambda$= 550nm. This figure shows all rocky planets detected with $R_\mathrm{p}<1.4R_\oplus$.
    Left panel shows {the measured quantity of} $\beta=A_\mathrm{g}R^2\Phi(\alpha)$ as a function of instellation. Individual components of albedo, radius, and phase angle {drawn from the \texttt{Bioverse} planet generation process including our injected trend} are shown in the right subplots. This survey yields 44 exoEarth candidates, and the distribution of $\beta$ shows a decrease at  instellations corresponding to the HZ, reflecting the injected trend in albedo (see upper right panel). }
    \label{ex_survey}
\end{figure}

With these changes in place, we can now use \texttt{Bioverse} to simulate the scientific outputs of proposed HWO designs and understand the detection biases introduced by the survey methods. 

In Figure \ref{ex_survey} we show the results of a simulated \texttt{Bioverse} survey for an example HWO design. This example calculation scheduled a blind survey, generated a sample of stars and planets, assigned their fundamental properties, and computed their observability, adding measurement uncertainties.
{The telescope used for this simulated exoEarth survey had an inscribed diameter of 8m,} a rough upper limit for the design trade space under consideration for HWO.
Its coronagraph has an inner working angle of 2$\lambda/D$ and an outer working angle of 30 $\lambda/D$, with a minimum achievable contrast of $\log(C)=-10.6$. It yielded a total of 44 exoEarth candidates, and many other rocky planets outside the HZ.  Figure \ref{ex_survey} plots the distribution of the properties of rocky planets ($R_\mathrm{p}<1.4R_\oplus)$ characterized in the survey as a function of instellation, $S_\mathrm{eff}$. Note that lower values of $S_\mathrm{eff}$ represent farther distances from the star and larger values represent closer distances. The {left} panel of this figure shows the distribution of the quantity $\beta=A_\mathrm{g}R^2\Phi(\alpha)$, which is directly measurable by direct imaging observations. Ideally one would like to measure the geometric albedo of the planet directly, but without a full spectral retrieval and multi-epoch observations, albedo, radius and phase will remain degenerate. However, using our forward-models in \texttt{Bioverse} we can obtain the values of these parameters and investigate how they contribute to the value of $\beta$, even if they aren't directly measurable by themselves (see the right panels of the plot).

Here we have injected the step function trend in albedo described earlier to planets in our dataset. This is shown in the upper right panel of Figure \ref{ex_survey} and is reflected in the distribution of $\beta$ as a dip at fluxes corresponding to the HZ. The distributions of radius and phase angle reproduce our intuition about the biases of the direct imaging method -- that brighter planets are easier to detect. We can see that at further distances from the star (smaller $S_\mathrm{eff}$), there is a bias towards detecting larger planets at fuller phases as one might expect since such planets would be brighter in reflected light. 

It should be noted that there is inherent stochasticity in the results from the random planet generation in \texttt{Bioverse}. Different surveys with the same input parameters may result in stronger or weaker evidence of a trend based on the random scattering of points. Therefore, it is necessary to replicate surveys such as that depicted in Figure \ref{ex_survey} many times in order to determine the recoverability of injected trends.


\subsection{Survey Distributions}
\label{survey_dists_section}

\begin{figure}
    \centering
    \plotone{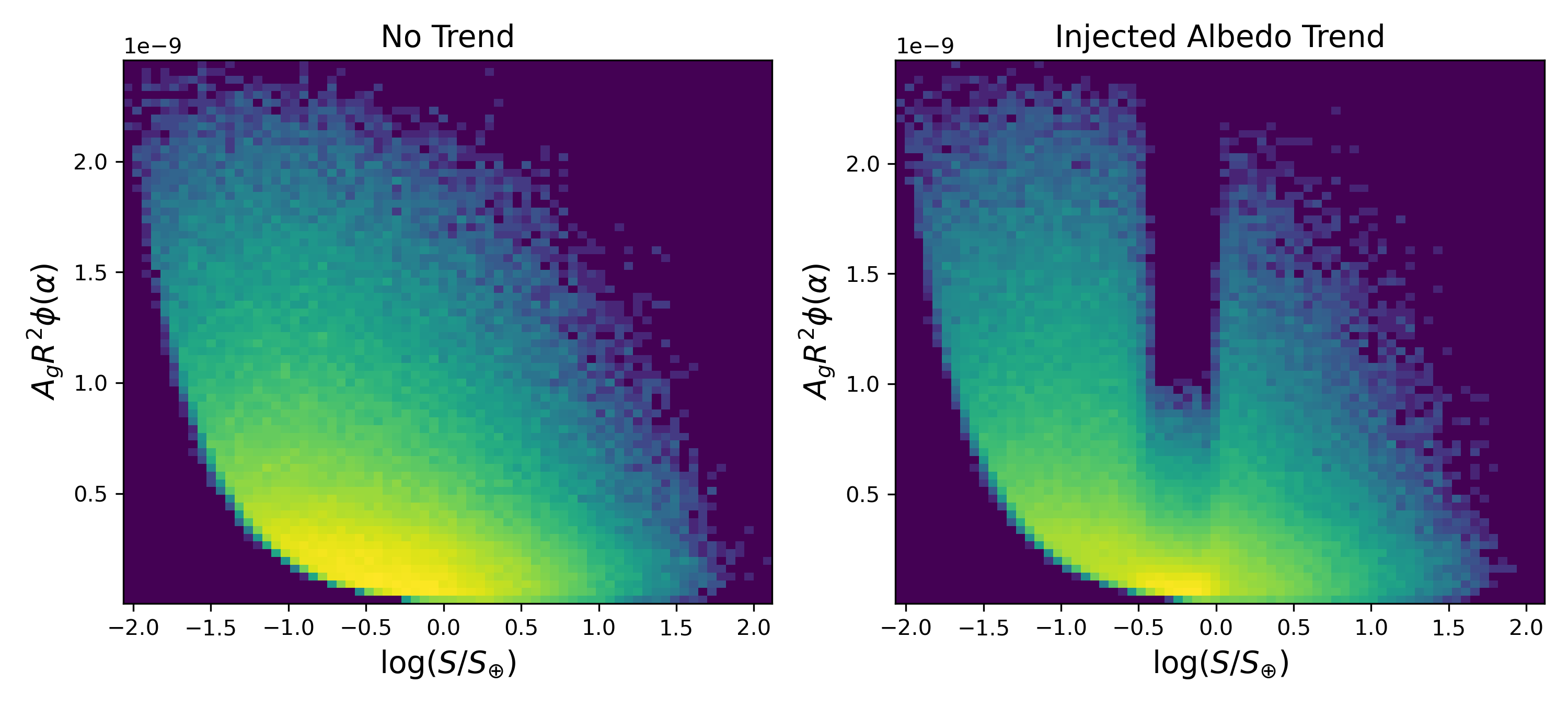}
    \caption{Survey distribution of rocky planets that would be detectable for a space-based direct imaging survey with an 8m diameter telescope.  Left panel shows the distribution of planets if there is no injected trend in albedo, while the right panel shows the case where there is a strong trend in albedo ($\Delta A = 0.4$). This distribution was generated by simulating the output of exoEarth surveys (analogous to Figure \ref{ex_survey}) for 1000 simulated universes. {Plots are color-coded based on the number of recovered planets in each bin}.  One can clearly see the influence of population-level trends in albedo in the output, where there is a prominent dip at instellations corresponding to the HZ.}
    \label{survey_dist}
\end{figure}

{We are} primarily concerned with how the sample size of exoEarth candidates affects the recoverability of trends in albedo, and our methodology focuses on constraining this sample size. Individual \texttt{Bioverse} simulations result in different numbers of exoEarths detected per survey, so {to understand} how a given sample size affects our ability to test hypotheses, we need to modify the standard procedure used by \texttt{Bioverse}.

As mentioned earlier, \texttt{Bioverse} consists of three modules on planet generation, survey simulation, and hypothesis testing \citep{Bixel2021}. For each simulation, \texttt{Bioverse} generates its planets, calculates their observability and computes the Bayesian evidence in favor of a trend, repeating this simulation multiple times to constrain the statistical power in favor of a hypothesis. 
We slightly modify the baseline procedure of \texttt{Bioverse} for our study. Rather than running all three modules together for each iteration, we instead run the planet generation and survey simulation modules separately from the hypothesis testing module. For a large number of simulated universes, we generate random planets and simulate their observability with a given mission design, combining the results of all these simulations to forward-model the distribution of observable planets that such a mission would be sensitive to. We refer to this as the \textit{survey distribution}. 
We then draw random samples from the survey distribution with a specified number of EECs and compute the Bayesian evidence for a trend. We continue drawing samples until we can constrain the statistical power to which a trend can be detected.

Figure \ref{survey_dist} shows the survey distribution for \texttt{Bioverse} simulations with the same inputs as those used in the example simulation shown in Figure \ref{ex_survey}. These simulations model how a strong trend in albedo as a function of instellation would appear in the distribution of the observables from a direct imaging survey with an 8m HWO design. In this figure, we combine the results of 1000 simulations to construct a survey distribution of detectable planets. The left panel of the figure shows the distribution when there are no injected trends while the right panel shows the scenario when there is a strong albedo trend.
In these distributions{,} one can see a strong edge in the lower left due to the minimum achievable contrast of the coronagraph. The main difference between these two distributions is that, for the case of an injected trend, there is a clear dip in the quantity $\beta$ around the HZ.  This implies that the non-detection of planets in this region of parameter space will be key in differentiating between these two scenarios.

The survey distribution shown here was constructed using simulations assuming an 8m diameter telescope. In Appendix \ref{sensitivity_appendix} we investigate how modifying mission design such as the telescope inscribed diameter and coronagraph minimum achievable contrast affect the survey distribution.
Ultimately we find that the shape of the survey distribution is remarkably insensitive to the mission architecture. 

\subsection{Statistical Analysis}
Using the survey distribution for an 8m LUVOIR-B like design for HWO, we draw random samples of a given number of EECs. To ensure that these samples contain planets outside the HZ as well, we sample random planets from the survey distribution until we reach a given number of EECs.

The statistical problem we are faced with in this study is as follows. Suppose we have a random sample of points such as those in the scatter plot in Figure \ref{ex_survey}. We want to determine whether these points were drawn from either of the distributions in Figure \ref{survey_dist}, i.e. whether they are consistent with the distributions with or without an injected trend in albedo. 
For multidimensional datasets such as ours, it is not straightforward to determine whether a sample of points is consistent with a given distribution. There are well established statistical tests in the one dimensional case, such as the Anderson-Darling test and KS test, but these cannot easily be generalized to a multidimensional case \citep{Rizzo2019}.

In our analysis we take a Bayesian approach to model comparison. We have two models which we would like to compare. We shall refer to the model with an injected trend as ``Trend" and the model without a trend as ``Null". We refer to our dataset of points as ``Data". Ultimately we want the ratio of probabilities that these models are true given the data. After applying Bayes Rule we get: 
\begin{equation}
    \frac{P(\mathrm{Trend|Data})}{P(\mathrm{Null|Data})}=\frac{P(\mathrm{Data|Trend})}{P(\mathrm{Data|Null})}\frac{P(\mathrm{Trend})}{P(\mathrm{Null})}
\end{equation}
If both models are assumed to have equal prior probabilities ($P(\mathrm{Trend}) = P(\mathrm{Null})$), then the main quantity we are concerned with is the Bayes Factor, $K= \frac{P(\mathrm{Data|Trend})}{P(\mathrm{Data|Null})}$. 

To compute the Bayes factor we need to know the marginal likelihoods in favor of each hypothesis, $P(\mathrm{Data|Model})$. We compute the marginal likelihoods using the survey distributions generated for both models. We use a kernel density estimate to approximate the survey distribution as a normalized probability density function {(see Appendix \ref{kde_section})} . For each point in the dataset, $ x_i \in \mathrm{Data}$, we calculate the probability of that point using the kernel density function and take the product of all data points. This gives 
\begin{equation} \label{evidence_eq}
    \log P(\mathrm{Data|Model})=\sum^{n}_{i=0} \log P(x_i|\mathrm{Model})
\end{equation}

We note that the Bayes factor is calculated using the marginal likelihoods for each model. For a parametric model this would require one to marginalize over all values of the model parameters. However, in our case we are dealing with a non-parametric dataset. We want to know whether our data is consistent with one set of points or another, not whether a given parametric model can fit the dataset. In this case there is no parameter to marginalize over, so the marginal likelihood is just given by the likelihood calculated using the kernel density estimate of the survey distribution.

Using this methodology, we can calculate the relative probabilities that a dataset was drawn from a distribution with an injected albedo trend or without a trend. We can interpret values of the Bayes Factor using the {Jeffreys' Scale} where values of $K>3$ are considered substantial evidence in favor of a trend, and values of $K>10$ constitute strong evidence \citep{Jeffreys1998}.

\section{Results}

\begin{figure}
    \centering
    \plotone{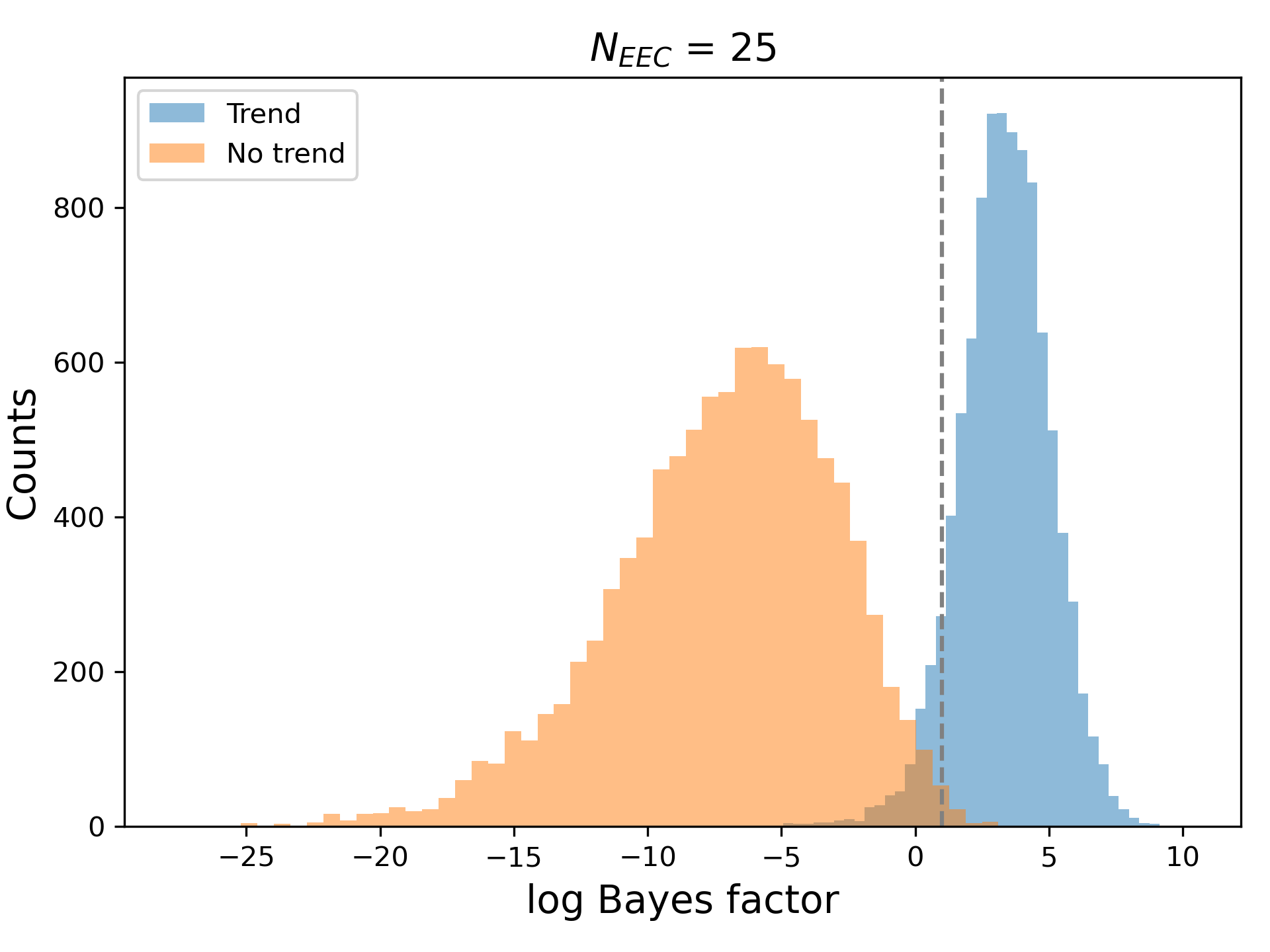}
    \caption{Histogram of Bayes factors for random samples with 25 exoEarth candidates (EECs). Blue histogram represents points drawn from a distribution with a strong injected trend in albedo ($\Delta A=0.4$). Orange histogram represents points drawn from a distribution without any trend in albedo. The dashed vertical line represents the threshold for strong detection of a trend at a Bayes factor of K=10. The region of the blue histogram to the right of this threshold represents the true positive rate, or statistical power in favor of the albedo trend model. The region of the orange histogram to the right of the threshold represents the false positive rate.}
    \label{bayes_factor_hist}
\end{figure}

\subsection{Calculating Statistical Power}
\label{example_analysis}

We apply the statistical methodology described in the previous section to an example scenario where we have a strong trend in albedo with $\Delta A=0.4$ and $A_0=0.7$ outside the HZ. We sample $N_{EEC}=25$ exoEarths alongside all other planet types from the survey distribution shown in Figure \ref{survey_dist}. For this sample of EECs we calculate the Bayes factor, taking the ratio of marginal likelihoods that the data are consistent with a distribution with or without a trend in albedo.   

Note that there is a large degree of stochasticity between random samples of the same sample size. Some will appear to present clear evidence in favor of a trend while others will be very ambiguous. This will be reflected by a large spread in Bayes factors between random samples. To account for this randomness, we simulate the results of a large number of simulations and find the fraction of simulations where the Bayes factor exceeds a threshold value for strong evidence of a trend (such as $K >10$). In Appendix \ref{n_sim_appendix} we determine how many simulations are needed for this analysis, settling on a value of around 1000 - 10,000 simulations.

For a set of 10,000 samples of 25 exoEarths from the survey distribution, we plot a histogram of the values of the Bayes factor in Figure \ref{bayes_factor_hist}. The blue histogram shows the Bayes factors for samples drawn from a survey distribution with a strong albedo trend. The dashed vertical line represents our threshold for statistical significance at $K=10$. The region of the blue histogram to the right of this threshold gives the true positive rate. This is the fraction of simulations where an albedo trend is detected given that the dataset actually has an injected trend. It can be interpreted as the statistical power in favor of the hypothesis, and in this case the value is $P(K>10|\mathrm{True})=0.921$. If we use a more optimistic criterion for detection of the trend of $K>3$, then the statistical power is $P(K>3|\mathrm{True})=0.954$. The orange histogram represents the Bayes factors for random samples drawn from a survey distribution {when} there is no injected trend in albedo. Most points in this distribution are at very small values of the Bayes factor, where there is strong evidence in favor of the no trend model over the albedo trend model. However there is a small tail of the distribution that exceeds the significance threshold. This represents the false positive rate, where samples with no injected trend in albedo are found to exhibit evidence in favor of a trend. For {a} threshold value of K=10 we find the false positive rate is $P(K>10|\mathrm{False})=0.0052$, but if we relax the threshold to be K=3  it increases to $P(K>3|\mathrm{False})=0.0105$.

\subsection{Varying Sample Size}
\label{vary_sample_size}

\begin{figure}
    \centering
    \plotone{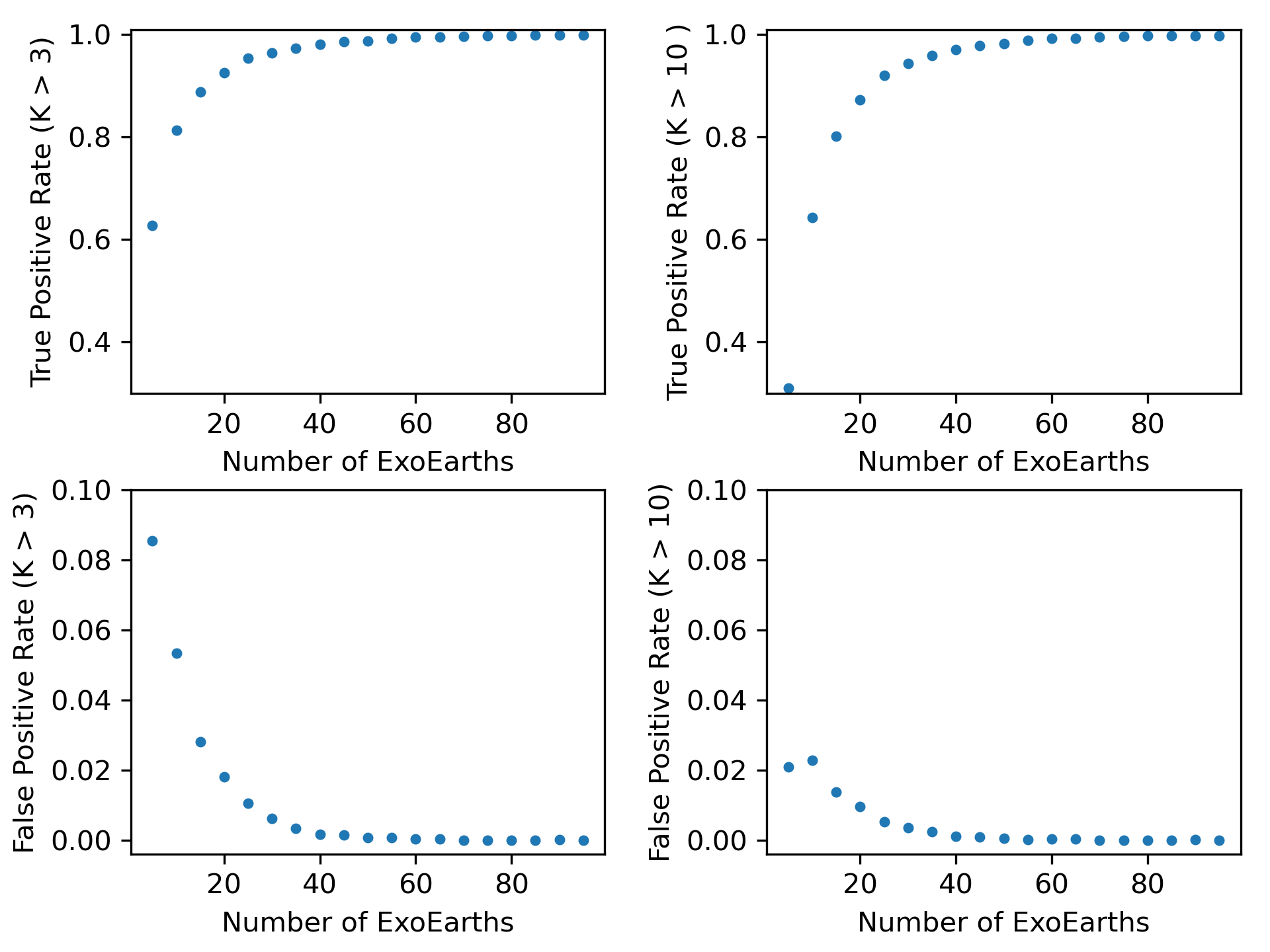}
    \caption{Sample size of exoEarths required to detect an injected trend in albedo. Top panels represent the true positive rate or statistical power as a function of the sample size of exoEarths, $N_{EEC}$, for two threshold values of the Bayes factor, $K$. Bottom subplots show the false positive rate as a function of sample size. For each point in these subplots an analysis similar to that in Figure \ref{bayes_factor_hist} was performed using 10,000 simulations of planetary samples at that sample size. Simulations injected a strong albedo trend  with $\Delta A = 0.4$.}
    \label{sample_size_analysis}
\end{figure}

We expand on the analysis from the previous section by allowing the number of EECs in the sample to vary as a free parameter. For an array of EEC sample sizes, we repeat this analysis, determining the statistical power and false positive rate for each value of $N_{EEC}$. This will enable us to determine the required sample size to recover the injected trend in albedo.

Figure \ref{sample_size_analysis} shows the effects of changing $N_{EEC}$ on our ability to recover albedo trends. The top panels show how the true positive rate, or statistical power varies as a function of $N_{EEC}$. This is done for two threshold values of the Bayes factor. The threshold value of K=10 gives a more conservative estimate, where stronger evidence is required to confidently detect a trend. For that threshold, in order to obtain a statistical power of 95\% one requires a sample size of around 30-35 EECs.
The bottom subplots show the false positive rates for these two thresholds. The plot of $P(K>3|\mathrm{False})$ gives a more conservative estimate of the false positive rate where less strong evidence of a trend will still count as a false positive. For this quantity, the false positive rate drops to 1\% at a sample size of around 30 EECs.

\subsection{Varying Trend Strength}

\begin{figure}
    \centering
    \plottwo{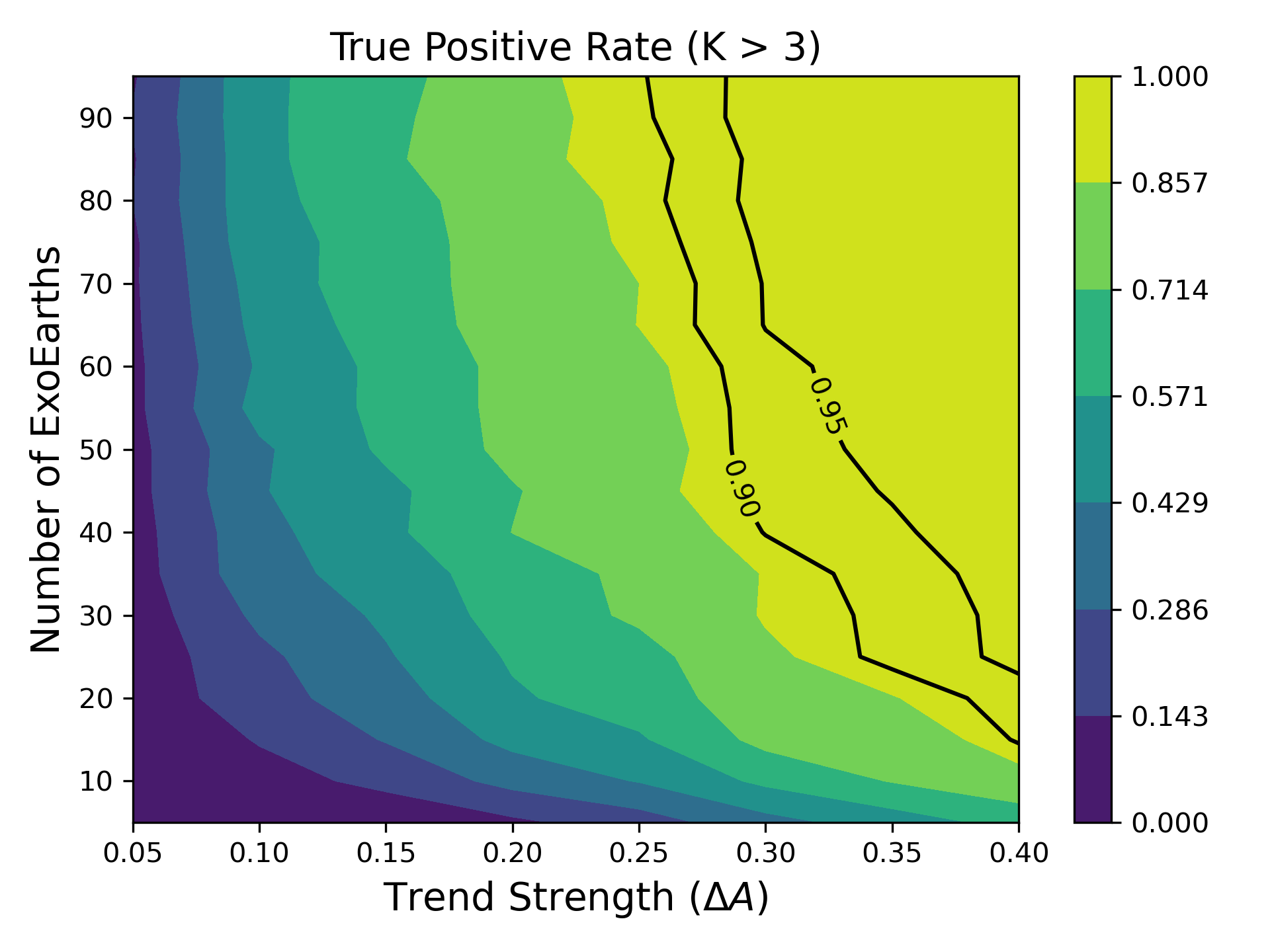}{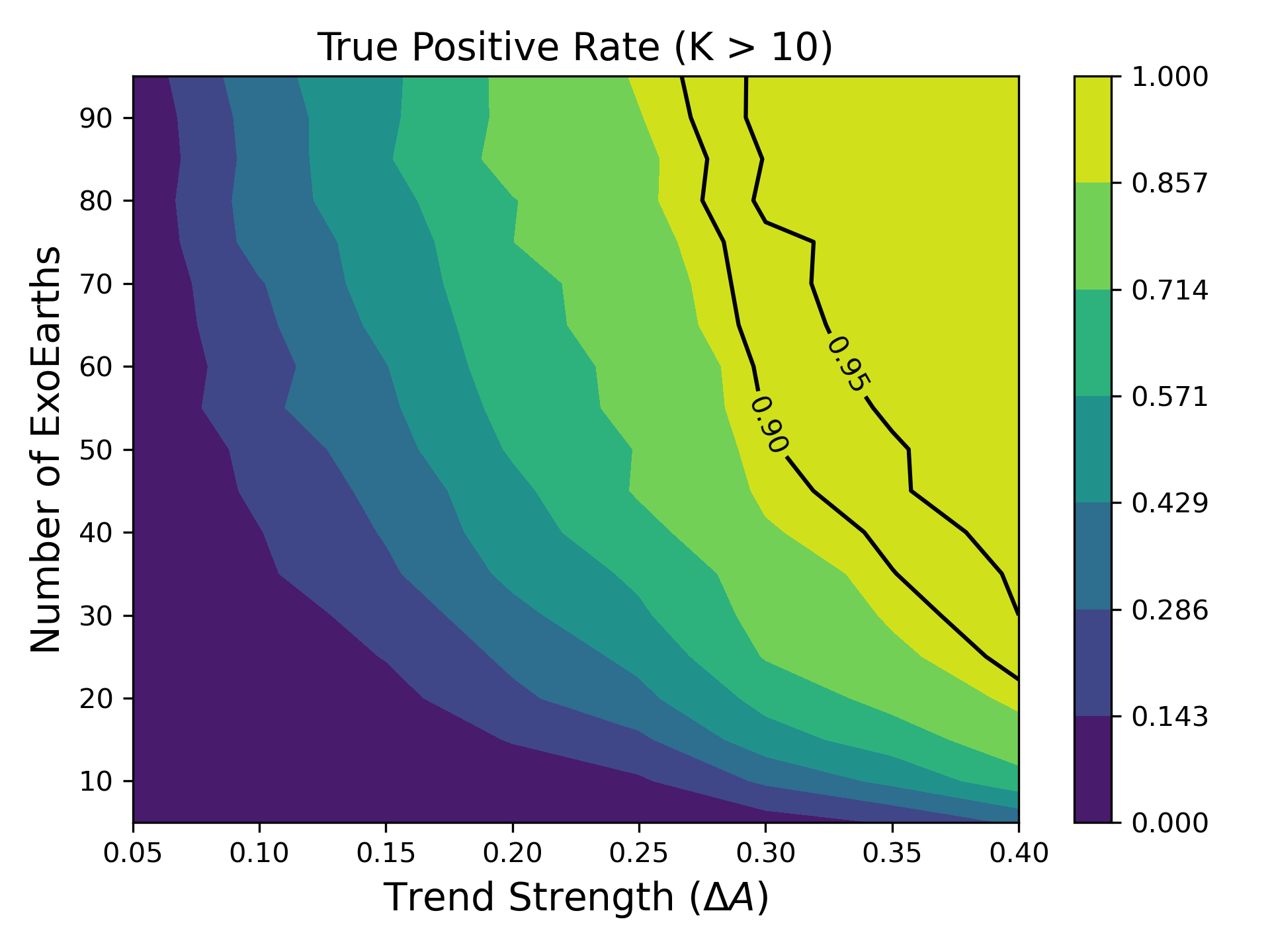}
    \caption{True positive rate as a function of trend strength, $\Delta A$, and sample size, $N_{EEC}$. Left and right subplots show results for two different threshold values for the Bayes factor required to detect an injected trend in albedo. Contours are shown for 90\% and 95\% statistical power. One can observe that the sample size required to detect a trend increases rapidly as the strength of the trend is reduced.}
    \label{true_positive_grid}
\end{figure}

The analyses in the previous sections focused on a specific strong trend in albedo with $A_0=0.7$ outside the HZ and $\Delta A=0.4$ in the HZ. In this section we determine how the sample size required to recover an albedo trend varies with the strength of the injected trend (parameterized by $\Delta A$). For an array of $\Delta A$ values between 0 and 0.4 we generate a survey distribution for planets generated with each trend. We then repeat the sample size analysis of Section \ref{vary_sample_size} for each value of $\Delta A$. This gives a grid of points in $\Delta A$ and $N_{EEC}$ where each point constitutes an analysis in the form of that in Section \ref{example_analysis}, with 1000 simulations to constrain the true and false positive rates.

This grid of points is plotted as a contour plot in Figure \ref{true_positive_grid}. It shows the statistical power that one could detect an injected trend of a given strength with a given sample size. Intuitively, in this plot the true positive rate increases as the strength of the trend or the number of EECs increases. For the subplots representing different threshold values of the Bayes factor we show contours of 90\% and 95\% true positive rates. One can observe that a relatively low sample size is required to recover strong trends at high statistical power. However, as the strength of the trend decreases, the number of EECs required to confidently detect it increases rapidly. At a trend strength of $\Delta A=0.40$ one requires a sample of around 30 EECs to achieve a statistical power of 95\% (using a threshold of $K>10$, see left panel). However, if the strength of the trend is decreased to $\Delta A=0.3$, the sample size required to achieve 95\% statistical power increases to 80-90 EECs. 
{Such a sample size is not expected to be easily achievable with a 6m diameter telescope, though simulations of very ambitious design scenarios sometimes achieve comparable yields when accounting for the large uncertainty in survey parameters such as $\eta_\oplus$ and the amount of exozodiacal dust per star (see long tails of the distribution of exoEarth yields in Figure 18 of \citealt{Stark2024}).  }
The required sample sizes are somewhat smaller when one uses a more optimistic threshold value for the Bayes factor required to detect a trend, but the general shape of the contour plot remains the same. 
In these plots there is some roughness in the contours. This is caused primarily by the coarseness of the grid points in $N_{EEC}$ and $\Delta A$, and also by the finite number of simulations. The grid resolution and number of simulations could potentially be increased, but at a large computational cost that is not justified by the slight improvement precision.

\begin{figure}
    \centering
    \plottwo{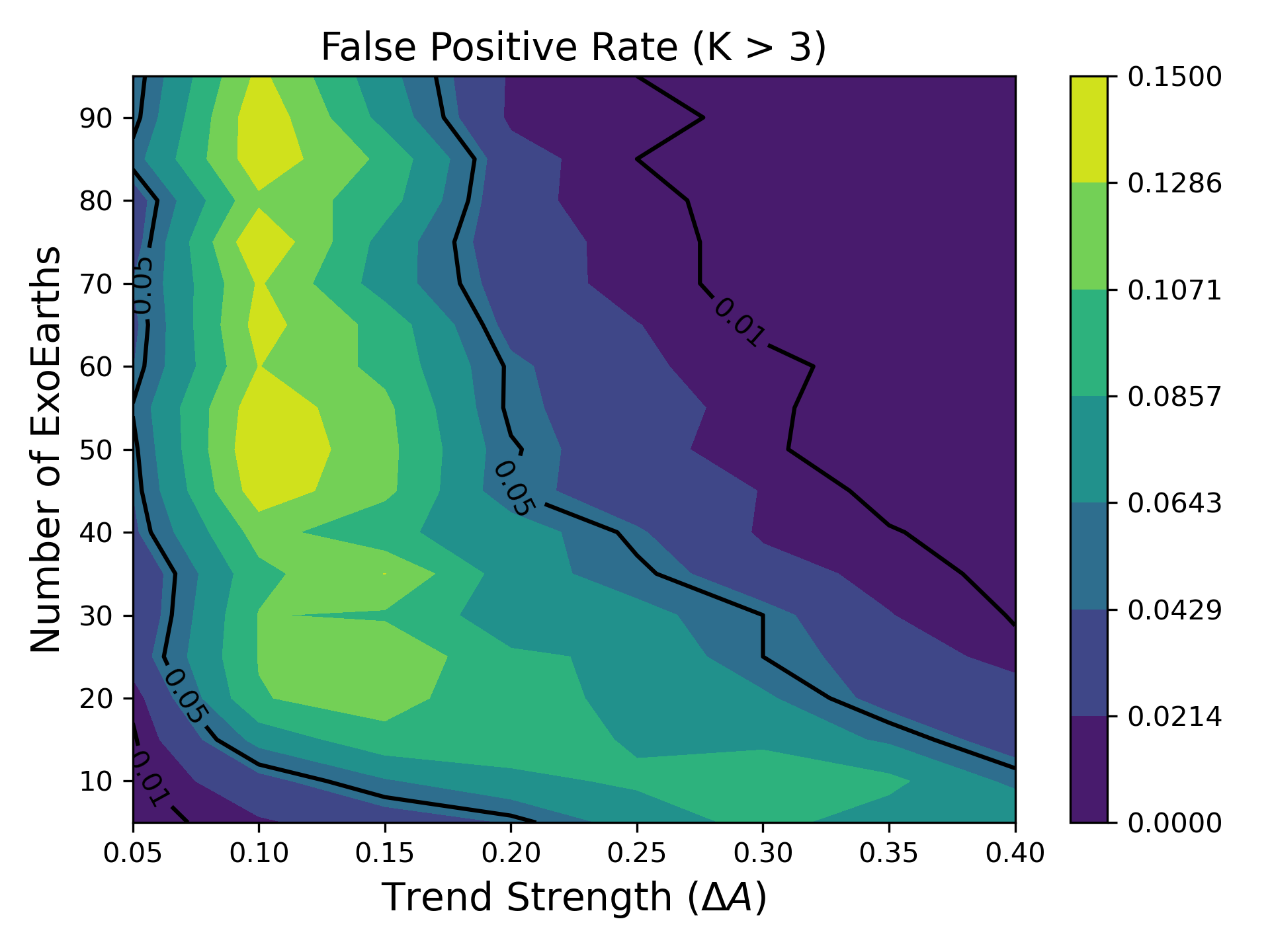}{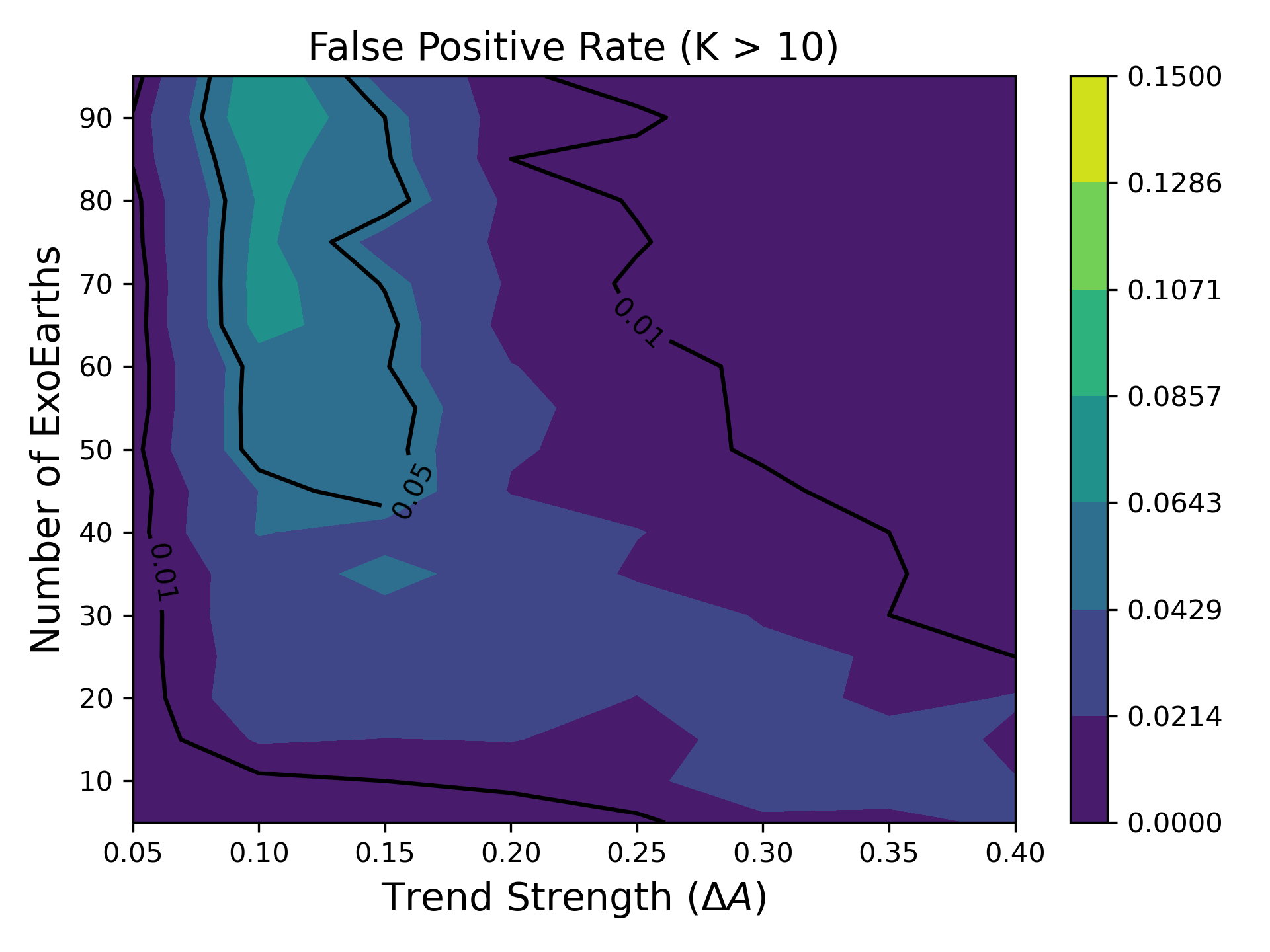}
    \caption{False positive rate as a function of trend strength and sample size. Left and right subplots show results for two different threshold values for the Bayes factor required to detect an injected trend in albedo. Contours are shown for a 1\% and 5\% false positive rate. Weaker trends and smaller samples of exoEarths have higher false positive rates.}
    \label{false_positive_grid}
\end{figure}

Figure \ref{false_positive_grid} shows how the false positive rate changes as a function of trend strength and sample size. We plot contours for 1\% and 5\% false positive rates. As noted earlier is a large amount of roughness in these plots due to the relatively small number of simulations. Computing the false positive rate requires one to accurately integrate over the upper tail of the distribution of Bayes factors, as seen in the orange histogram in Figure \ref{bayes_factor_hist}. This requires a large number of simulations to accurately characterize. With our current grid having 1000 simulations per grid point, we are able to constrain this quantity, but we should expect a significant degree of uncertainty for each point, especially for the higher $K>10$ threshold plot.
Despite these uncertainties we are able to characterize how the false positive rate changes with trend strength and sample size. We find that, in the region where statistical power is greater than 90\%, the false positive rate is unlikely to exceed 1\%. Higher false positive rates are achieved when the trend strength or sample size are reduced. There are also high false positive rates in the region of parameter space with low trend strength and high sample size. In the bottom left corner of the plots, the false positive rate appears to increase as trend strength and $N_{EEC}$ increase. Looking at the individual distributions of Bayes factors calculated from samples without an injected trend (orange histogram in Fig \ref{bayes_factor_hist}), we observe that this appears to be due to a real phenomenon. As sample size or trend strength increases, the mean of the Bayes factor distribution moves to smaller values as one might expect, but the width of the distribution also increases so that a larger portion of the distribution is above the significance threshold. The exact reason for the increase in the variance of the Bayes factor distribution remains unclear, but we can see that the shape of the contour plot in Figure \ref{false_positive_grid} is influenced by this behavior.  Ultimately these plots suggest that a survey would be most strongly limited by obtaining a detection of a trend at high statistical power rather than by the false positive rate. The false positive rate only has meaningfully large values in regions of parameter space where the statistical power is very low, so one {would not} have a good chance of confidently detecting a trend anyways.

\section{Discussion}
\label{discussion_section}

In our analysis, we investigated the ability of future direct imaging missions to recover population-level trends in albedo from measurements of the directly observable properties of planet-star contrast and separation. 
Varying the strength of underlying albedo trends and the sample size of exoEarths in a survey, we determined the statistical power to which these trends could be recovered. 
{Simulating a space-based direct imaging survey, we found that to recover the strongest trends in albedo one requires a sample size on the order of the Decadal Survey's target of 25 exoEarths. However, for lower trend strengths the required sample size increases rapidly, implying that in order to detect such weak trends one would require an ambitious telescope design capable of surpassing the the EEC yield goals set by the Decadal Survey. 
The values we report for the number of exoEarths required to recover a trend in albedo are rough estimates. These estimates depend on assumptions about the nature of planetary systems such as the planetary distribution function and the amount of exozodiacal dust, which are highly uncertain (see next section). There is also a smaller degree of uncertainty introduced by elements of our analysis such as the choices involved in constructing the survey distribution (Appendix \ref{sensitivity_appendix}), the kernel density estimate used (Appendix \ref{kde_section}), and the significance threshold for the detection of a trend. Nonetheless, even if there is some uncertainty present in the absolute numerical figures presented in this study, it is able to accurately determine how our ability to recover population-level trends in albedo depends on the sample size of exoEarths and the strength of the trend, allowing us to assess the feasibility of studying this science case with future direct imaging missions.}

We hope to be able to uncover the distribution of rocky planet albedos as a function of instellation.
While our simplest toy model for a strong step function trend in albedo could be recovered by HWO-like {mission} designs, the actual distribution of planetary albedos {likely has} a more complicated and subtle dependence on instellation. With sample sizes of around 25 EECs, it is unlikely that one would be able to differentiate between models of the albedo distribution that don't have major differences between them. Furthermore, the strength of the trend in albedo as a function of instellation is likely to be much weaker than the strongest step functions we tested.
It is reasonable to expect that a significant fraction of rocky planets in the HZ may not exhibit habitable conditions. The classical model for the HZ assumes that an Earth-like $\mathrm{CO}_2$ cycle helps to regulate the planet's climate \citep{Kasting1993hz}. This would require the presence of plate tectonics, but we currently lack knowledge as to whether plate tectonics are a common phenomenon for Earth-sized planets, or whether the Earth is unusual in this regard \citep{Affholder2025}. If a large portion of rocky planets in the HZ are not Earth-like, this would dampen the strength of the observable trend in albedo, making it much more difficult to recover.

While there are definitely obstacles to recovering trends in albedo for rocky planets, it does not seem to be unachievable. 
There are many potential ways in which one could improve the ability of a direct imaging survey to recover population-level trends in albedo. Here we will detail a few of them.

\subsection{Decreasing noise}

\begin{figure}
    \centering
    \plotone{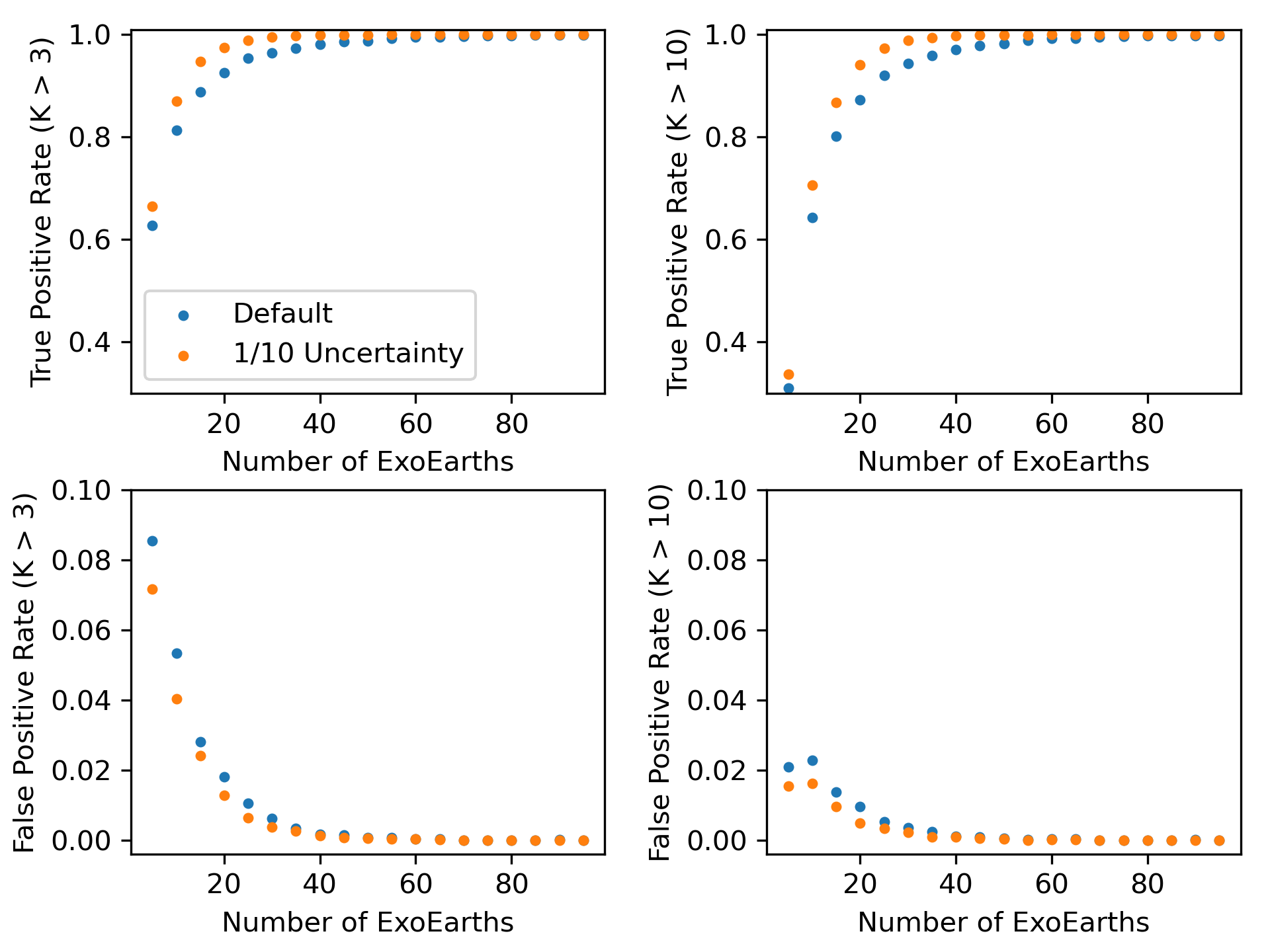}
    \caption{Same as Figure \ref{sample_size_analysis} but demonstrating the effects of lowering the uncertainty in observed properties. Blue points are from simulations with standard uncertainty levels, while orange points have uncertainty in contrast and {semi-major} axis reduced by a factor of 10.}
    \label{low_noise}
\end{figure}

One potential way to improve our ability to recover albedo trends is by improving the precision of measurements. As described in Section \ref{noise_section} we have injected realistic levels of measurement noise in our simulations. This has the effect of increasing the difficulty of recovering a trend when compared to noiseless models. If we reduce the simulated noise levels, we can reveal the effect of obtaining more accurate measurements.
While precision in luminosity will have a limit imposed by model uncertainties and systematics, planet-star contrast and {semi-major} axis can be readily measured to higher precision. Increasing the exposure time per target will increase the signal to noise, resulting in lower contrast uncertainties. Similarly, increasing the number of observational epochs would allow one to constrain a planet's {semi-major} axis to greater precision. Modifying a survey to increase the precision of these properties would likely have the effect of decreasing the overall survey yield, as more time would need to be devoted to each individual target.

Suppose that we were somehow able to reduce the uncertainties in contrast and {semi-major} axis by a factor of 10. 
Figure \ref{low_noise} illustrates how decreasing the measurement noise in contrast and {semi-major} axis affects one's ability to recover an albedo trend. The trend we consider here is a strong step function in albedo ($\Delta A=0.4$), identical to that shown in Figure \ref{sample_size_analysis}. Points with standard uncertainties are shown in blue, while points with uncertainties reduced by a factor of 10 are shown in orange. One can see that decreasing the uncertainties increases the statistical power to which a trend can be detected and reduces the false positive rate. This decreases the sample size of EECs required to obtain 95\% statistical power, but the improvement is relatively minor and likely would not justify the tradeoffs.

{Contrast and semi-major axis are two of the observables that we have the most control over when designing a survey, but there are many more factors that affect our ability to recover trends in albedo. One key factor is the planet distribution function, specifically for rocky planets in the habitable zone. In our study, we used the SAG 13 occurrence rates, which roughly extrapolate the planet distribution function to encompass Earth-like radii and orbital periods. Our knowledge of planet occurrence rates in this region of parameter space is highly uncertain, but the results of this study are largely independent of the assumed value for $\eta_\oplus$, as we determined the sample sizes needed to recover albedo trends rather than the yield of planets from a given mission design. However, the shape of the planetary distribution function has the potential to influence our analysis, as changes to the distribution function used to generate synthetic planets will propagate to affect our forward-modeled survey distributions. Increased constraints on the planetary distribution function could greatly improve the accuracy of our predictions, though we are unlikely to obtain many rocky planets in the habitable zones of FGK-type stars prior to HWO.}

{Another major factor that could improve our ability to estimate the detectability of albedo trends is our knowledge about the amount of exozodiacal dust per star (exozodi). 
To detect and characterize planets, we will have to estimate and subtract exozodi from images. The degree to which our models do not perfectly subtract exozodi will affect our estimates of planet flux, and increase the uncertainty in properties such as albedo, radius and phase function which influence the observed flux. The effects of exozodi on signal-to-noise and thus the contrast uncertainties are included in our exposure time calculator, but exozodi will also affect the uncertainties in other measured and estimated parameters. For instance, imperfect exozodi subtraction result in lower signal-to-noise, making it more difficult for spectral retrievals to constrain planetary fundamental properties and atmospheric compositions \citep{Salvador2024}. This in turn makes it more challenging to break the degeneracy between radius, albedo, and phase function.
In our calculations, we assumed a fixed level of exozodi per star (3 times the zodiacal light of the solar system, informed by the HOSTS survey of \citealt{Ertel2020}). However, for most of the target stars that could be observed by a survey, exozodi remain poorly constrained or lack measurements entirely. Future precursor observations will be necessary to estimate the levels of exozodi for promising target stars, ensuring that they are actually amenable to direct imaging observations. While not directly relating to our inferences of the required sample size of exoEarths, perhaps the largest effect exozodi will have is their effect on survey yield \citep{Stark2024}.
Higher exozodi levels will result in fewer confident planet detections at the required signal-to-noise. While the results of our analysis for the required sample size of exoEarths will not change, higher exozodi levels mean that a mission would have a more difficult time reaching this sample size.}

{Constraining our estimates of the planet distribution function and the number of exozodi per star is unlikely to decrease the sample size of exoEarths required to detect a trend in albedo. Instead it would increase the precision of our analysis, allowing us to obtain more accurate estimates of the required number of exoEarths to recover these trends.}

\subsection{Integrating color information}

So far our simulations have focused on monochromatic measurements in the visible bandpass. We have considered the planetary albedo only as it appears at 550 nm. However, HWO will have a wide spectral coverage ranging from the NIR to the UV. Even in the absence of full spectral retrievals, one could detect planets and measure their contrasts in the UV, optical, and NIR channels, providing rough color information about planet candidates. Past studies suggest that when observing planets in reflected starlight, color information can be vital to distinguish between different planet types \citep{Traub2003,Crow2011}. Planetary albedos can vary quite significantly as a function of wavelength and using albedo-color data would allow one to differentiate between planets with similar bulk properties such as super-Earths and sub-Neptunes. Color information could allow us to be more confident that the planets which we detect in the HZ are true exoEarth candidates rather than other planet types which could easily be confused as EECs \citep{Bixel2019}. {Combining both information about planetary colors and their albedo dependence with instellation has the potential to yield stronger constraints on the HZ than using both datasets separately. Understanding the dependence of albedo on wavelength and instellation will require additional theoretical work, but it may reduce the required sample size needed to empirically constrain the HZ.}

{We currently have a cursory understanding of how these albedo-instellation trends would appear in different bandpasses.  For our simplified model of planetary albedo at visible wavelengths (around V band centered at 549 nm), albedo was modeled as a step function in instellation in the HZ. Our choice to model a survey in the visible bandpass stems from the results of \citet{Stark2024b}, which suggest that the optimal wavelength to detect exoEarths with a LUVOIR-B-like mission would be around 500 nm. At other wavelengths, a blind survey is unlikely to return as high a yield of exoEarths, but we can still gain a heuristic understanding of the strength of an albedo-instellation trend based on the differences in albedo between Earth and other solar system bodies, as well as the albedos of modeled snowball Earth planets at the outer edge of the habitable zone. The albedo difference between Earth and Venus decreases towards shorter wavelengths in the near UV because Earth's albedo increases (until falling off rapidly after around 350 nm) while Venus's albedo decreases \citep{roberge2017}. In the near IR, Earth's albedo slightly increases while Venus's slightly decreases, also reducing the difference between them. One would expect similar behavior when comparing Earth to a globally glaciated planet with an Earth-like atmosphere \citep{Shields2013}. For such a planet, albedo would be expected to peak around 500 nm, decreasing sharply towards shorter wavelengths and gradually decreasing towards the IR. If this pattern holds, it suggests that the albedo difference between exoEarths and planets outside the HZ would be highest in the visible. At near UV wavelengths, one might expect the albedo difference to be much smaller, whereas in the near IR the trend strength would be dampened but not to the same extent. }

{This dependence of the albedo trends on wavelength is highly speculative and is informed primarily from solar system observations or planetary models computed in a different context. This highlights the need for additional theoretical work and planetary modeling in order to determine the expected relation between planetary albedo and instellation, and how this relation is dependent on wavelength. Such theoretical work falls outside the scope of this study, but we emphasize that with additional modeling results, one may be able to obtain constraints on the empirical boundaries of the HZ that would not be achievable using albedo-instellation relations or albedo-color trends alone.}

\subsection{Full spectral retrievals}

Our analysis thus far has focused on the observables available from planet \textit{detections} alone. These direct observables (such as contrast and separation) will be available for a much larger population of planets than the sample of planets with high enough signal to noise spectra for full spectral characterization.  If we instead focus on planets with fully characterized spectra, then a great deal more information becomes available such as {planetary} atmospheric compositions, and bulk properties \citep{Damiano2022}.  Observing weaker trends in albedo may require the additional constraints on planetary properties available from full spectral retrievals. 

Spectral characterization of directly imaged exoplanets will be a challenging task, requiring long exposure times to obtain relatively low resolution noisy spectra of planetary atmospheres. However, fitting atmospheric models to planetary spectra would likely greatly enhance our ability to detect trends in albedo, as they will enable us to place constraints on planetary radius and albedo \citep{Salvador2024}. Alongside multi-epoch measurements of planet-star separation to constrain illumination phases, spectral retrievals will allow us to break the apparent degeneracy between albedo, radius, and phase. Our analysis focused on measurements of the quantity $\beta=A_\mathrm{g}R_\mathrm{p}^2\Phi(\alpha)$ which is related to planetary luminosity and embodies this degeneracy between properties. However if planet phase and radius can be inferred independently, then we would be able to obtain direct estimates of the planet's albedo by itself. This would allow us to place constraints on the underlying distribution of albedo as a function of $S_\mathrm{eff}$ without the effects of other confounding variables. 

It remains an open question as to whether a survey focusing on spectrally characterized exoEarths would would improve our ability to identify population-level trends in albedo. On one hand, data from spectral retrievals would allow for direct estimation of planetary geometric albedos. These will probably be at a relatively low precision, but will enable one to determine whether trends in contrast are actually due to albedo rather than the confounding factors of radius and phase. On the other hand, the sample size of planets with fully characterized spectra will be much smaller than the entire population of planets observed by a mission such as HWO. Our analysis in this study highlighted that the sample size of EECs observed by a direct imaging mission will pose a major limit on our ability to infer trends about the planetary population. As higher signal-to-noise spectra are time consuming to obtain, this will reduce the sample size of planets which it is possible to fully characterize.
Whether the benefits of increased fidelity measurements of planetary albedo will outweigh the costs of a reduced sample size is not clear and will require additional research.

Beyond the specific hypothesis addressed in this study, full spectral retrievals would also allow one to reveal trends in planetary composition that could provide alternative means to empirically constrain the boundaries of the HZ. In addition to planetary colors addressed earlier, previous studies have considered possible methods to constrain the HZ based on planetary compositions. For example, \citet{Bixel2019} investigated the hypothesis that the HZ would host a larger fraction of planets with detectable water vapor in their atmospheres. Alternatively one could measure $\mathrm{CO}_2$ concentrations in the atmospheres of planets and determine whether they are consistent with the theoretical predictions of the habitable zone in terms of the dependence of $\mathrm{CO}_2$ on $S_\mathrm{eff}$ based in the silicate weathering feedback cycle {\citep{Bean2017,Checlair2019,Turbet2020}}. We intend to apply our updated \texttt{Bioverse} model for direct imaging to revisit these tests of the HZ in future work.  Taking the union of several independent tests of the concept of the HZ may allow one to place the strongest constraints on the empirical HZ boundaries. 

We should note that in order to identify empirical evidence of the HZ in planetary population, it is not enough to only characterize the Earth-sized planets we find in the HZ. One would also require full spectral retrievals for rocky planets outside the HZ. As seen in our earlier analysis, the presence of rocky planets outside the HZ is critical for detecting any trend in albedo in the HZ. In order to determine whether planets in the HZ are quantifiably different from those outside of it, one needs to observe both planets inside and outside the HZ. These concerns may influence the survey strategy employed by HWO if one's goal is to better understand rocky planets as a population. Perhaps focusing our efforts towards only HZ planets will have a negative impact on mission science return, biasing our survey based on our limited current day understanding of habitability.

\section{Conclusions}

Future space-based direct imaging missions such as the Habitable Worlds Observatory aim to discover Earth-sized planets in the HZ. We hope to study {exoEarths} as a population and uncover the physical processes that influence their habitability, formation, and evolution. However it is unclear whether {these} missions will be able to obtain a large enough sample of exoEarths to make inferences about the population. Our primary focus was thus to identify the sample size of planets required to uncover trends in the exoEarth population and whether this sample size would be achievable for currently considered mission designs. 

In this study, we focused {primarily} on population-level trends in albedo which could be {measured by direct imaging detections alone} without the need for full spectral characterization. {Such trends could provide empirical constraints on the boundaries of the HZ due to the fact that} the classical picture of the HZ predicts that rocky planets in the HZ will have lower albedos than those outside of it. Using a toy model for albedo vs instellation, we investigated how strong a trend had to be to be recoverable and what sample sizes were required.  We adapted the \texttt{Bioverse} statistical planetology code to simulate an HWO-like direct imaging survey to detect exoEarths. Generating a synthetic population of planets, we injected trends in albedo and simulated the detectability of these planets and the statistical power to which we could recover the injected trends. We simulated an 8m HWO architecture, but we found that our predictions for the required sample size of exoEarths were largely independent of assumed mission design (See discussion in Appendix \ref{sensitivity_appendix}). This implies that our findings are generalizable for direct imaging mission designs spanning the currently considered HWO design trade space.

We found that strong trends in {exoEarth albedos} ($\Delta A=0.4$) could be readily detected by a space-based direct imaging mission. One can recover this trend with a statistical power of 95\% given a sample size of around 30 exoEarths. This sample size is potentially achievable with 6m diameter HWO architectures given various possible design trades (see for example the yield calculations of \citet{Stark2024}). Reducing the strength of the injected trend caused the required sample size to increase rapidly. This suggests that HWO or similar missions will be able to detect strong trends in albedo in the exoEarth population, but subtler trends may require unfeasibly large sample sizes to recover with a high level of confidence.
To improve the ability of direct imaging surveys to recover these population-level trends and place empirical constraints on the HZ, we discussed several possible strategies in Section \ref{discussion_section}. These include improving the measurement precision for contrast and separation, incorporating additional observables such as planet colors, and using direct constraints on albedo from full spectral retrievals. All of these approaches have potential benefits and tradeoffs. Most prominently, for the case of full spectral retrievals one obtains stronger constraints on albedo at the cost of a reduced sample size.

{Ultimately, constraining the boundaries of the habitable zone using planetary albedos is on the edge of feasibility using near-term space-based direct imaging. In this study, we demonstrated the ability of the \texttt{Bioverse} framework to simulate the ability of future direct imaging missions to test hypotheses such as this albedo-instellation relation. In future research we intend to apply the methodology developed in this study to investigate the detectability of other population-level trends using future telescopes such as HWO. Using a performance metric of the number of exoEarths required to test these hypotheses, future trade studies will be able to determine whether specific HWO designs will be able to meet this target sample size, assessing whether or not a given science case is feasible.}

\section*{Acknowledgments}
We thank Ravi Kopparapu for his insights into the trends in albedos that would be expected from the classical model of the HZ.
We acknowledge Steve Bryson and Eric Feigelson for providing consultation on the statistical methods used in this study.
NWT is supported by an appointment with the NASA Postdoctoral Program at the NASA Goddard Space Flight Center, administered by Oak Ridge Associated Universities under contract with NASA.
The results reported herein benefited from collaborations and/or information exchange within NASA’s Nexus for Exoplanet System Science (NExSS) research coordination network sponsored by NASA’s Science Mission Directorate under Agreement No. 80NSSC21K0593 for the program ``Alien Earths".

\appendix

\section{Survey Distribution Sensitivity}
\label{sensitivity_appendix}

\begin{figure}
    \centering
    \plotone{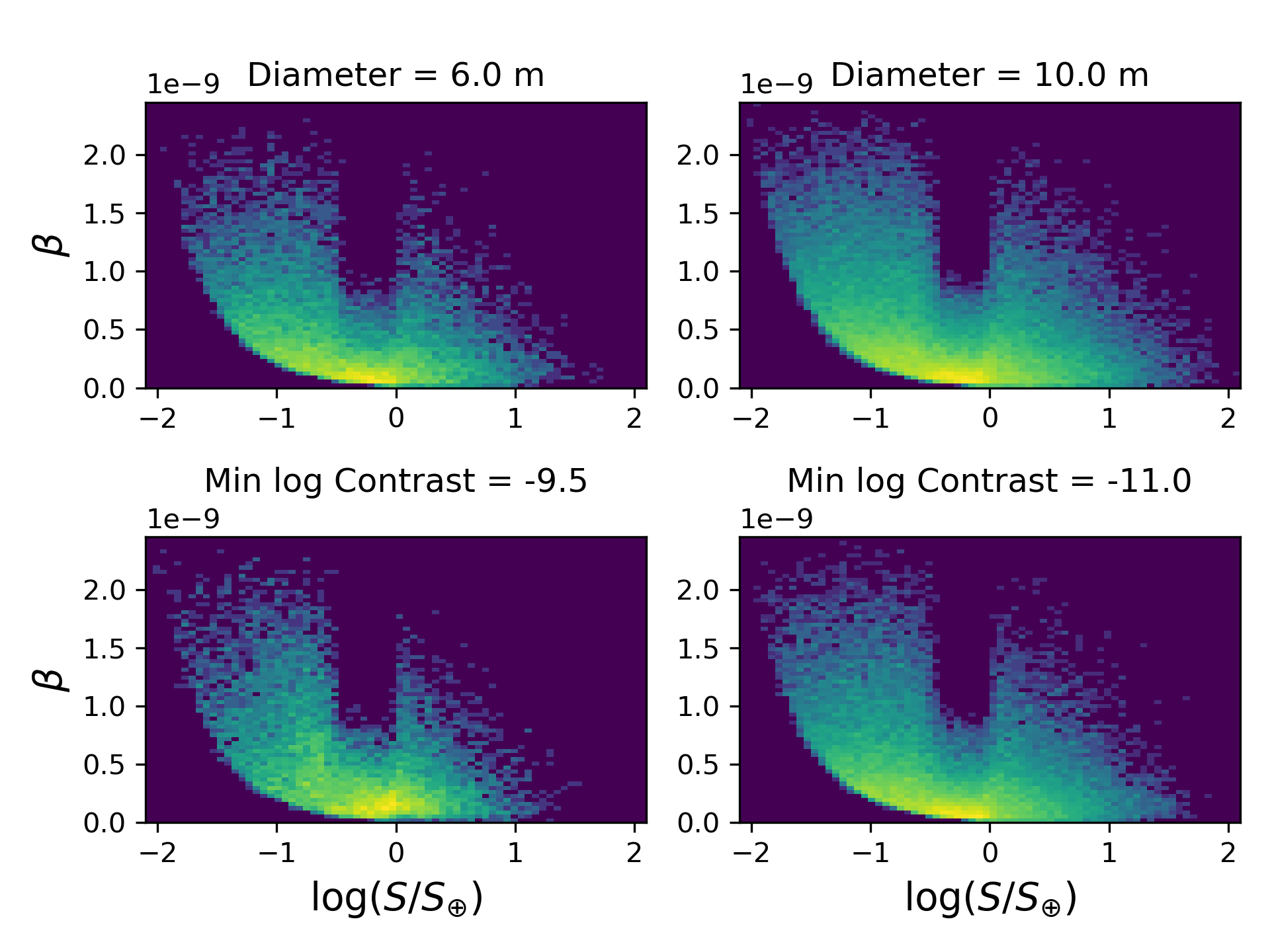}
    \caption{Insensitivity of the survey distribution on telescope and coronagraph parameters. Each of these scenarios take the baseline case of the survey distribution shown in Figure \ref{survey_dist} with D = 8m and $\log C_{min} = -10.6$, varying either the diameter or minimum achievable contrast. Despite second order differences between these scenarios, they largely have the same overall shape of their survey distribution.}
    \label{survey_dist_sensitivity}
\end{figure}

The survey distributions used in this study assume an {HWO-like} telescope and coronagraph design described in section \ref{survey_sim}. However, we would like for our results to be independent of assumptions about mission architecture. We aim to estimate the sample size of exoEarths required to recover the injected trend in albedo, but we don't want such estimates to be sensitive to assumed parameters such as the telescope diameter used to generate the survey distribution.
Here we briefly discuss how changing the design of the telescope and coronagraph affect the survey distribution. Figure \ref{survey_dist_sensitivity} shows how the survey distribution changes as a function of telescope and coronagraph design. We can see that even when one varies the telescope diameter between 6 and 10 meters the underlying distribution of planets is largely the same. There are a few smaller differences between the distributions. The 6m distribution is composed of much fewer points overall as each simulated survey has a much lower yield than the 10m case. There is also a slight difference at higher instellations as the 10m telescope is able to detect planets at a closer angular separation to the star because the coronagraph IWA scales as $\lambda/D$. We also investigate the effects of changing the minimum contrast the coronagraph can achieve, analogous to the noise floor. We find that modifying the minimum contrast has a slight effect on the cutoff curve in the bottom left of the distribution, but the most prominent effect is that much fewer planets are detected if a survey cannot achieve deep contrast ratios. 

Ultimately the survey distribution appears to be very insensitive to any differences in telescope and coronagraph design. The differences between survey distributions with different mission architectures are second order effects and the overall shape of the distribution remains nearly the same. It is very unlikely that one could distinguish between samples of points drawn from each of these distributions. For the purposes of our study we are able to say confidently that a sample drawn from the survey distribution of an 8m telescope will be largely consistent with that of a 6m or 10m telescope, so that we can use that specific survey distribution to simulate sample sizes that may not be achieved unless one goes to smaller or larger telescope diameters. There is a small degree of uncertainty introduced in drawing from a survey distribution with a fixed telescope and coronagraph design, but it appears to be smaller than other uncertainties implicit in survey simulation, such as planet occurrence rates and the amount of exozodiacal light per star. 

The similarity in shape of these survey distributions suggest that the sample size of EECs required to confidently recover trends in albedo shouldn't change much between different proposed HWO designs, although whether or not those proposed designs will be able to meet that yield target is another question. This suggests that the results from our analysis should be generalizable to other mission architectures under consideration.

\section{Kernel Density Estimates}
\label{kde_section}

\begin{figure}
    \centering
    \plotone{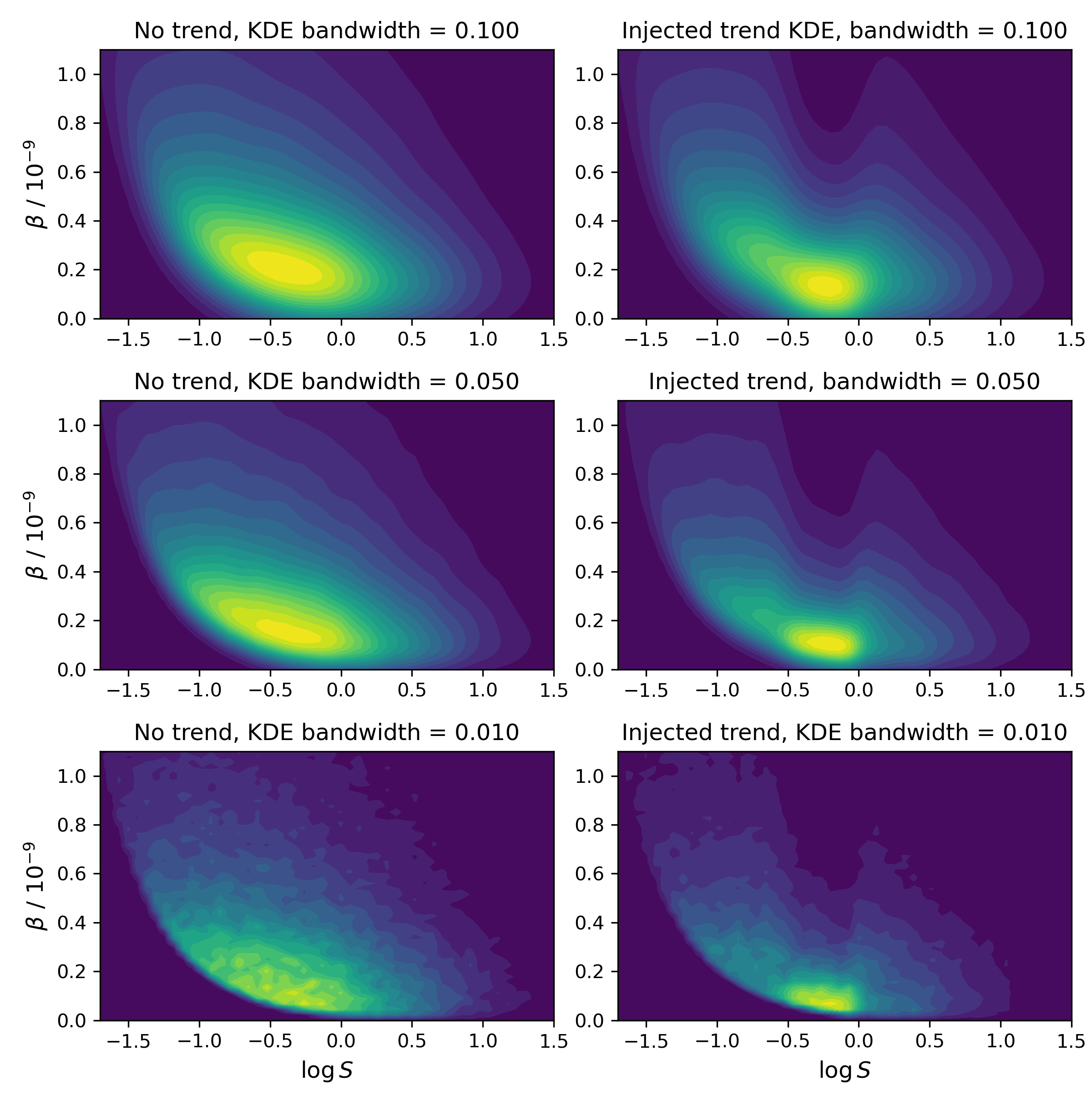}
    \caption{{Comparison of kernel density estimates of the survey distribution. Left panels show a survey distribution with no injected trend while right panels show a distribution with a strong injected trend ($\Delta A=0.4$). These kernel density estimates used a Gaussian kernel, changing the bandwidth parameter in each row.} }
    \label{kernel_compare}
\end{figure}

\begin{figure}
    \centering
    \plotone{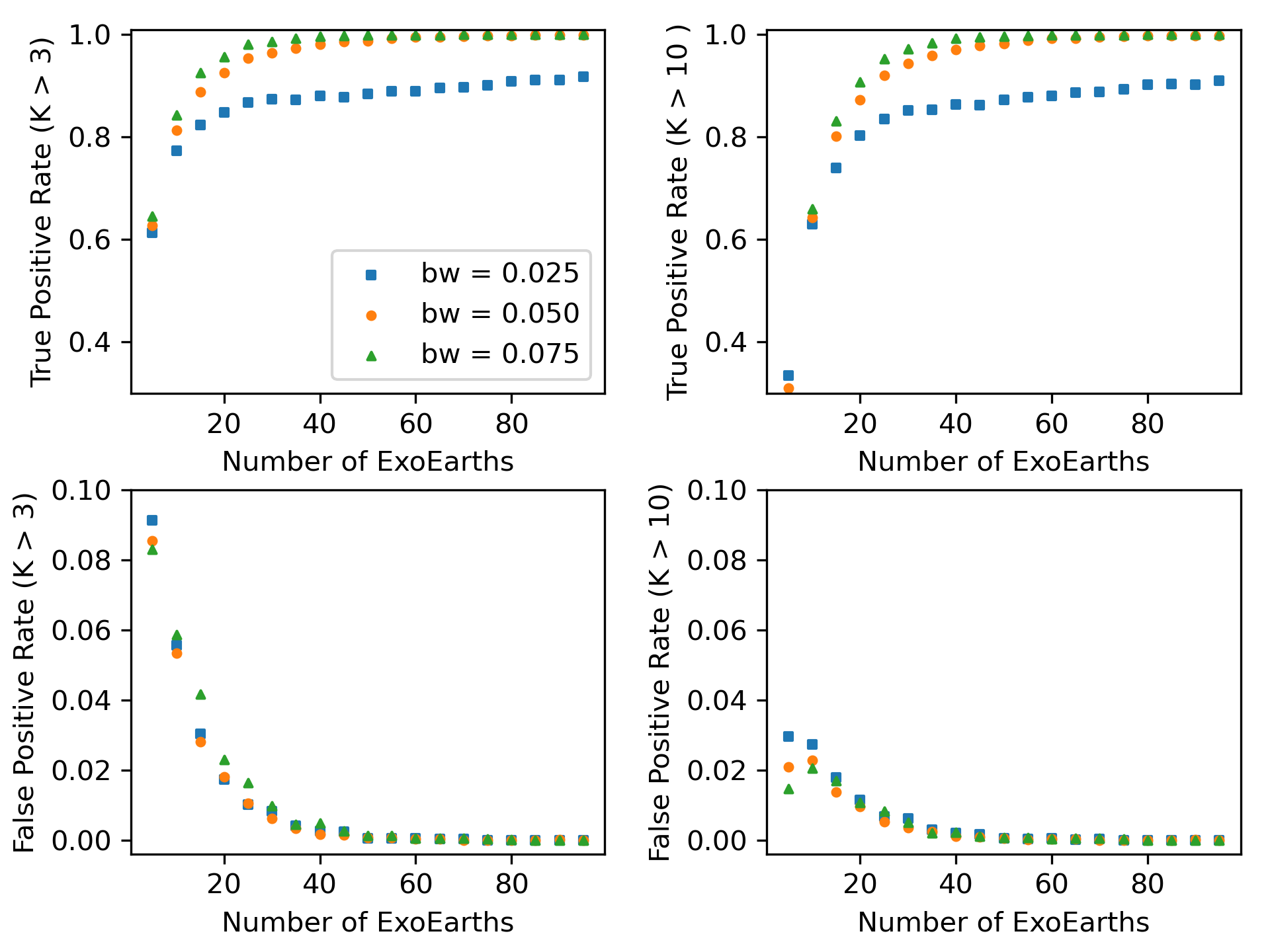}
    \caption{{Sensitivity of the results of our analysis to the choice of kernel for the kernel density estimate. Simulations are the same as those in Figure \ref{sample_size_analysis}, but change the bandwidth parameter (bw) of the Gaussian kernel used.}}
    \label{kernel_sensitivity}
\end{figure}

{For our forward-modeled ``survey distributions" described in Section \ref{survey_dists_section}, we use a kernel density estimate (KDE) to approximate the discrete data points as a smooth, normalized probability density function. This KDE is used to compute the marginal likelihood that the results of a simulated survey were drawn from this underlying distribution. Before applying our KDE, we first preprocessed our data, dividing $\beta$ by a factor of $10^{-9}$ and taking $\log_{10}$ of $S_\mathrm{eff}$ to get both axes to have similar orders of magnitude.
We used a Gaussian kernel for our KDEs with a bandwidth parameter of 0.05. We selected this value for kernel bandwidth as we found that it captured the most of the underlying structure of the survey distribution without emulating the sampling noise in the training dataset or smoothing over key features. Figure \ref{kernel_compare} illustrates that too large of a bandwidth parameter results in overly smooth KDEs, while too small a bandwidth results in a large amount of noise.}

{Selecting a KDE that accurately represents the survey distribution is important as the marginal likelihoods used in our analysis are sensitive to the choice of kernel used in the KDE. This is because the marginal likelihoods are given by the product of probabilities for each data point from a simulated survey (see Eq \ref{evidence_eq}), so small changes in these individual probabilities can propagate to become large differences in the product. 
To determine how robust the results of our analysis are to changes in the kernel, we repeat part of our analysis of the sample size required to detect strong ($\Delta A =0.4$) trends in albedo, as described in Section \ref{vary_sample_size}, varying the kernel used in the KDE. In this analysis, we considered 50\% variations of the Gaussian kernel bandwidth around our selected value of 0.050. We can observe in Figure \ref{kernel_sensitivity} that decreasing the kernel bandwidth by 50\%  noticeably lowers the inferred true positive rate of a survey (statistical power), while increasing it by 50\% only results in very small increase in the true positive rate. The dependence of true positive rate on the sample size of exoEarths appears to be similar between the simulations with different kernels, though absolute values are somewhat different and the lower bandwidth case appears like it may asymptote at a value less than 100\% statistical power. Interestingly, the false positive rates seem to be largely unaffected by kernel bandwidth, with only small differences around the smallest sample sizes.}

{Our results for true positive rates appear much more sensitive to lowering the bandwidth than to increasing it. This behavior appears to be due to our choice of a kernel bandwidth of 0.050 which was the near the cutoff value of bandwidth beyond which the KDE becomes too noisy. Increasing the bandwidth slightly just results in a slightly more smoothed distribution while decreasing the bandwidth results in a much noisier KDE. 
In the smaller bandwidth case, neither of the KDEs with or without an injected trend fit are a good match to simulated datasets, so their marginal likelihoods are much smaller. This in turn affects the Bayes factors, resulting in lower Bayes factors on average and thus lower statistical power.}

{The results of this analysis highlight the importance of selecting a KDE that accurately represents the structure of the survey distributions. Our choice of kernel reasonably preserves intrinsic structure of the survey distribution without overfitting the the sampling noise, but the values reported in this study for statistical power and the required number of planets should be taken with a grain of salt due to their potential sensitivity on KDE kernels. There may be potential methods to decrease the sensitivity of these results on one's choice of kernel.  Simulating more data points in the survey distribution could make the training dataset have more complete coverage of the parameter space and make density estimates less noisy. However, the current method we employ to calculate the KDE uses the Euclidean distance between a sampled point and all training data points in the dataset. Increasing the number of points in this dataset would increase the computation time required for this analysis to the point where it may no longer be tractable. One could attempt to address this issue by using a different algorithm for the distance metric, but this would cause difficulty in obtaining an accurate normalization for the KDE. Mitigating the sensitivity of our results to the choice of KDE kernel choice may not be the optimal use of resources as it is potentially dwarfed by the effects of astrophysical uncertainties such as our lack of knowledge of the planetary distribution function in the HZ and the exozodiacal light per star. }

\section{Constraining the required number of simulations}
\label{n_sim_appendix}
\begin{figure}
    \centering
    \plotone{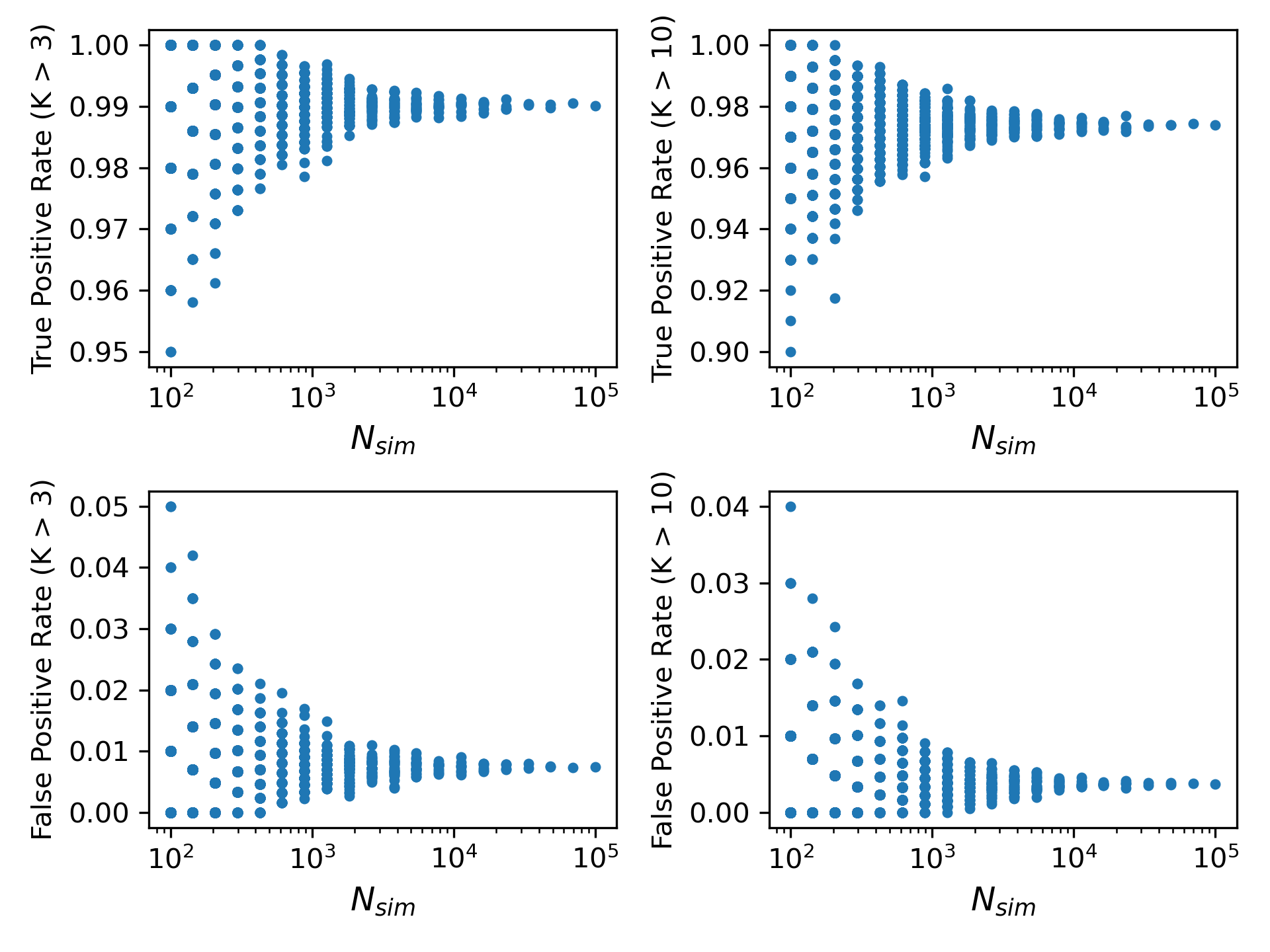}
    \caption{Required number of simulations to constrain the true positive rates (top panels) and false positive rates (bottom panels). Sub-samples of size $N_{sim}$ are drawn from a sample of 100,000 simulations, and true and false positive rates are calculated for each sub-sample. One can observe high variability for sub-samples with low $N_{sim}$, but as $N_{sim}$ increases the calculated quantities converge to a given value. These simulations were from a scenario with $\Delta A=0.4$ and $N_{EEC}=25$. Left and right columns represent different threshold values for the Bayes factor required for detection of a trend.}
    \label{req_num_sims}
\end{figure}

In our statistical analysis, we want to obtain accurate values for the true positive rate, $P(K>10|\mathrm{True})$,   and false positive rate, $P(K>10|\mathrm{False})$. For a given trend strength and sample size, generating the results of a single simulation from the survey distribution results in a single value for the Bayes factor. We repeat the sample generation process several times to populate a histogram of Bayes factors analogous to Figure \ref{bayes_factor_hist}. We aim to constrain the fraction of simulations that exceed a given threshold value of the Bayes factor, allowing us to calculate the true and false positive rates. To accurately predict these quantities we need to know how many simulations are required. In particular the false positive rate probes the extreme tails of the orange distribution in Figure \ref{bayes_factor_hist} so very large numbers of simulations may be required.

To address the number of simulations needed, we observed how true and false positive rates varied as a function of the number of simulations used to calculate them. For a given scenario ($\Delta A=0.4$, $N_{EEC}=25$) we ran 100,000 simulations obtaining the same number of Bayes factors. Such a large number of simulations would be computationally intractable to perform for a large number of scenarios, but for this specific scenario we ran a large number of simulations to determine how many simulations were actually needed for our analysis. From this large sample of simulations we can draw sub-samples with a specified number of simulations and calculate the true and false positive rates for each one. We do this for values of the number of simulations ranging from 100 to 100,000. This is shown in Figure \ref{req_num_sims}, where each calculated quantity is shown in a different subplot. We can observe how the calculated values of true and false positive rates vary for each value of $N_{sim}$, the number of simulations. One can see that there {are} relatively large uncertainties {for} small numbers of simulations, but as $N_{sim}$ increases, the estimates of true and false positive rates converge to a given value. Note that the fraction of simulations with a Bayes factor exceeding a given value can only take discrete values, as it is a finite number of simulations divided by the total number of simulations. This behavior is seen most prominently for the lowest numbers of simulations. From the results of this analysis it seems that a value of $N_{sim}=1000$ is sufficient to reduce the uncertainty in the true positive rate to around 1-2\%. False positive rates are more difficult to constrain. Ideally, around 10,000 simulations would be useful to reduce their uncertainties, but this number of simulations may not be feasible for a large multidimensional analysis. A value of $N_{sim}=1000$ may be able to place sufficient constraints on false positive rates, even if they will be a bit noisy. Note that this analysis on the required number of simulations was performed for a single scenario, and the $N_{sim}$ requirements may change as a function of the strength of injected trends and the number of EECs.

\bibliographystyle{aasjournal}
\bibliography{sources}

\begin{thebibliography}{}
\expandafter\ifx\csname natexlab\endcsname\relax\def\natexlab#1{#1}\fi
\providecommand{\url}[1]{\href{#1}{#1}}
\providecommand{\dodoi}[1]{doi:~\href{http://doi.org/#1}{\nolinkurl{#1}}}
\providecommand{\doeprint}[1]{\href{http://ascl.net/#1}{\nolinkurl{http://ascl.net/#1}}}
\providecommand{\doarXiv}[1]{\href{https://arxiv.org/abs/#1}{\nolinkurl{https://arxiv.org/abs/#1}}}

\bibitem[{{Affholder} {et~al.}(2025){Affholder}, {Mazevet}, {Sauterey}, {Apai}, \& {Ferri{\`e}re}}]{Affholder2025}
{Affholder}, A., {Mazevet}, S., {Sauterey}, B., {Apai}, D., \& {Ferri{\`e}re}, R. 2025, \aj, 169, 125, \dodoi{10.3847/1538-3881/ada384}

\bibitem[{{Apai} {et~al.}(2017){Apai}, {Cowan}, {Kopparapu}, {Kasper}, {Hu}, {Morley}, {Fujii}, {Kane}, {Maley}, {del Genio}, {Karalidi}, {Komacek}, {Mamajek}, {Mandell}, {Domagal-Goldman}, {Barman}, {Boss}, {Breckinridge}, {Crossfield}, {Danchi}, {Ford}, {Iro}, {Kasting}, {Lowrance}, {Madhusudhan}, {McElwain}, {Moore}, {Pascucci}, {Plavchan}, {Roberge}, {Schneider}, {Showman}, \& {Turnbull}}]{Apai2017}
{Apai}, D., {Cowan}, N., {Kopparapu}, R., {et~al.} 2017, arXiv e-prints, arXiv:1708.02821, \dodoi{10.48550/arXiv.1708.02821}

\bibitem[{{Apai} {et~al.}(2019){Apai}, {Banzatti}, {Ballering}, {Bergin}, {Bixel}, {Birnstiel}, {Bose}, {Brittain}, {Cadillo-Quiroz}, {Carrera}, {Ciesla}, {Close}, {Desch}, {Dong}, {Dressing}, {Fernandes}, {France}, {Gharib-Nezhad}, {Haghighipour}, {Hartnett}, {Hasegawa}, {Jang-Condell}, {Kalas}, {Kane}, {Kim}, {Krijt}, {Lisse}, {L{\'o}pez-Morales}, {Malhotra}, {Morrison}, {Mulders}, {Pontoppidan}, {Scharf}, {Schwarz}, {Schwieterman}, {Stassun}, {Turner}, {Wagner}, \& {Young}}]{Apai2019}
{Apai}, D., {Banzatti}, A., {Ballering}, N.~P., {et~al.} 2019, \baas, 51, 475

\bibitem[{Bean {et~al.}(2017)Bean, Abbot, \& Kempton}]{Bean2017}
Bean, J.~L., Abbot, D.~S., \& Kempton, E. M.-R. 2017, The Astrophysical Journal Letters, 841, L24

\bibitem[{Bixel \& Apai(2019)}]{Bixel2019}
Bixel, A., \& Apai, D. 2019, The Astronomical Journal, 159, 3

\bibitem[{Bixel \& Apai(2021)}]{Bixel2021}
---. 2021, The Astronomical Journal, 161, 228

\bibitem[{{Bryson} {et~al.}(2021){Bryson}, {Kunimoto}, {Kopparapu}, {Coughlin}, {Borucki}, {Koch}, {Aguirre}, {Allen}, {Barentsen}, {Batalha}, {Berger}, {Boss}, {Buchhave}, {Burke}, {Caldwell}, {Campbell}, {Catanzarite}, {Chandrasekaran}, {Chaplin}, {Christiansen}, {Christensen-Dalsgaard}, {Ciardi}, {Clarke}, {Cochran}, {Dotson}, {Doyle}, {Duarte}, {Dunham}, {Dupree}, {Endl}, {Fanson}, {Ford}, {Fujieh}, {Gautier}, {Geary}, {Gilliland}, {Girouard}, {Gould}, {Haas}, {Henze}, {Holman}, {Howard}, {Howell}, {Huber}, {Hunter}, {Jenkins}, {Kjeldsen}, {Kolodziejczak}, {Larson}, {Latham}, {Li}, {Mathur}, {Meibom}, {Middour}, {Morris}, {Morton}, {Mullally}, {Mullally}, {Pletcher}, {Prsa}, {Quinn}, {Quintana}, {Ragozzine}, {Ramirez}, {Sanderfer}, {Sasselov}, {Seader}, {Shabram}, {Shporer}, {Smith}, {Steffen}, {Still}, {Torres}, {Troeltzsch}, {Twicken}, {Uddin}, {Van Cleve}, {Voss}, {Weiss}, {Welsh}, {Wohler}, \& {Zamudio}}]{Bryson2021}
{Bryson}, S., {Kunimoto}, M., {Kopparapu}, R.~K., {et~al.} 2021, \aj, 161, 36, \dodoi{10.3847/1538-3881/abc418}

\bibitem[{Cahoy {et~al.}(2010)Cahoy, Marley, \& Fortney}]{Cahoy2010}
Cahoy, K.~L., Marley, M.~S., \& Fortney, J.~J. 2010, The Astrophysical Journal, 724, 189

\bibitem[{Checlair {et~al.}(2019)Checlair, Abbot, Webber, Feng, Bean, Schwieterman, Stark, Robinson, Kempton, Alcabes, {et~al.}}]{Checlair2019}
Checlair, J.~H., Abbot, D.~S., Webber, R.~J., {et~al.} 2019, arXiv preprint arXiv:1903.05211

\bibitem[{Crow {et~al.}(2011)Crow, McFadden, Robinson, Meadows, Livengood, Hewagama, Barry, Deming, Lisse, \& Wellnitz}]{Crow2011}
Crow, C.~A., McFadden, L., Robinson, T., {et~al.} 2011, The Astrophysical Journal, 729, 130

\bibitem[{{Damiano} \& {Hu}(2022)}]{Damiano2022}
{Damiano}, M., \& {Hu}, R. 2022, \aj, 163, 299, \dodoi{10.3847/1538-3881/ac6b97}

\bibitem[{Ertel {et~al.}(2020)Ertel, Defr{\`e}re, Hinz, Mennesson, Kennedy, Danchi, Gelino, Hill, Hoffmann, Mazoyer, {et~al.}}]{Ertel2020}
Ertel, S., Defr{\`e}re, D., Hinz, P., {et~al.} 2020, The Astronomical Journal, 159, 177

\bibitem[{Gaudi {et~al.}(2020)Gaudi, Seager, Mennesson, Kiessling, Warfield, Cahoy, Clarke, Domagal-Goldman, Feinberg, Guyon, {et~al.}}]{gaudi2020}
Gaudi, B.~S., Seager, S., Mennesson, B., {et~al.} 2020, arXiv preprint arXiv:2001.06683

\bibitem[{{Hardegree-Ullman} {et~al.}(2023){Hardegree-Ullman}, {Apai}, {Bergsten}, {Pascucci}, \& {L{\'o}pez-Morales}}]{HardegreeUllman2023}
{Hardegree-Ullman}, K.~K., {Apai}, D., {Bergsten}, G.~J., {Pascucci}, I., \& {L{\'o}pez-Morales}, M. 2023, \aj, 165, 267, \dodoi{10.3847/1538-3881/acd1ec}

\bibitem[{Hardegree-Ullman {et~al.}(2025)Hardegree-Ullman, Apai, Haffert, Schlecker, Kasper, Kammerer, \& Wagner}]{HardegreeUllman2025}
Hardegree-Ullman, K.~K., Apai, D., Haffert, S.~Y., {et~al.} 2025, The Astronomical Journal, 169, 171

\bibitem[{Jeffreys(1998)}]{Jeffreys1998}
Jeffreys, H. 1998, The theory of probability (OuP Oxford)

\bibitem[{Kasting {et~al.}(1993)Kasting, Whitmire, \& Reynolds}]{Kasting1993hz}
Kasting, J.~F., Whitmire, D.~P., \& Reynolds, R.~T. 1993, Icarus, 101, 108

\bibitem[{Kopparapu(2018)}]{Kopparapu2018review}
Kopparapu, R.~K. 2018, Handbook of Exoplanets, 58

\bibitem[{{Kopparapu} {et~al.}(2014){Kopparapu}, {Ramirez}, {SchottelKotte}, {Kasting}, {Domagal-Goldman}, \& {Eymet}}]{Kopparapu2014}
{Kopparapu}, R.~K., {Ramirez}, R.~M., {SchottelKotte}, J., {et~al.} 2014, \apjl, 787, L29, \dodoi{10.1088/2041-8205/787/2/L29}

\bibitem[{Kopparapu {et~al.}(2013)Kopparapu, Ramirez, Kasting, Eymet, Robinson, Mahadevan, Terrien, Domagal-Goldman, Meadows, \& Deshpande}]{Kopparapu2013}
Kopparapu, R.~K., Ramirez, R., Kasting, J.~F., {et~al.} 2013, The Astrophysical Journal, 765, 131

\bibitem[{Kopparapu {et~al.}(2018)Kopparapu, H{\'e}brard, Belikov, Batalha, Mulders, Stark, Teal, Domagal-Goldman, \& Mandell}]{Kopparapu2018}
Kopparapu, R.~K., H{\'e}brard, E., Belikov, R., {et~al.} 2018, The Astrophysical Journal, 856, 122

\bibitem[{Madhusudhan \& Burrows(2012)}]{madhusudhan2012}
Madhusudhan, N., \& Burrows, A. 2012, The Astrophysical Journal, 747, 25

\bibitem[{Mallama {et~al.}(2017)Mallama, Krobusek, \& Pavlov}]{mallama2017}
Mallama, A., Krobusek, B., \& Pavlov, H. 2017, Icarus, 282, 19

\bibitem[{Mamajek \& Stapelfeldt(2023)}]{mamajek2023}
Mamajek, E., \& Stapelfeldt, K. 2023, NASA ExEP Mission Star List for the Habitable Worlds Observatory: Most Accessible Targets to Survey for Potentially Habitable Exoplanets, \url{https://exoplanets.nasa.gov/internal_resources/2645_NASA_ExEP_Target_List_HWO_Documentation_2023.pdf}

\bibitem[{Morgan {et~al.}(2019)Morgan, Savransky, Turmon, Mennesson, Mamajek, Shaklan, Soto, Stapelfeldt, Dula, \& Keithly}]{morgan2019}
Morgan, R., Savransky, D., Turmon, M., {et~al.} 2019, in Techniques and instrumentation for detection of exoplanets IX, Vol. 11117, SPIE, 1111701

\bibitem[{{Mulders} {et~al.}(2015{\natexlab{a}}){Mulders}, {Pascucci}, \& {Apai}}]{Mulders2015a}
{Mulders}, G.~D., {Pascucci}, I., \& {Apai}, D. 2015{\natexlab{a}}, \apj, 798, 112, \dodoi{10.1088/0004-637X/798/2/112}

\bibitem[{{Mulders} {et~al.}(2015{\natexlab{b}}){Mulders}, {Pascucci}, \& {Apai}}]{Mulders2015b}
---. 2015{\natexlab{b}}, \apj, 814, 130, \dodoi{10.1088/0004-637X/814/2/130}

\bibitem[{National Academies~of Sciences \& Medicine(2021)}]{DecadalSurvey}
National Academies~of Sciences, E., \& Medicine. 2021, Pathways to Discovery in Astronomy and Astrophysics for the 2020s (Washington, DC: The National Academies Press), \dodoi{10.17226/26141}

\bibitem[{{Neil} \& {Rogers}(2020)}]{Neil2020}
{Neil}, A.~R., \& {Rogers}, L.~A. 2020, \apj, 891, 12, \dodoi{10.3847/1538-4357/ab6a92}

\bibitem[{{Pascucci} {et~al.}(2019){Pascucci}, {Mulders}, \& {Lopez}}]{Pascucci2019}
{Pascucci}, I., {Mulders}, G.~D., \& {Lopez}, E. 2019, \apjl, 883, L15, \dodoi{10.3847/2041-8213/ab3dac}

\bibitem[{Rizzo(2019)}]{Rizzo2019}
Rizzo, M.~L. 2019, Statistical computing with R (Chapman and Hall/CRC)

\bibitem[{Roberge {et~al.}(2017)Roberge, Rizzo, Lincowski, Arney, Stark, Robinson, Snyder, Pueyo, Zimmerman, Jansen, {et~al.}}]{roberge2017}
Roberge, A., Rizzo, M.~J., Lincowski, A.~P., {et~al.} 2017, Publications of the Astronomical Society of the Pacific, 129, 124401

\bibitem[{Robinson(2025)}]{Robinson2025}
Robinson, T.~D. 2025, arXiv preprint arXiv:2507.22258

\bibitem[{{Salvador} {et~al.}(2024){Salvador}, {Robinson}, {Fortney}, \& {Marley}}]{Salvador2024}
{Salvador}, A., {Robinson}, T.~D., {Fortney}, J.~J., \& {Marley}, M.~S. 2024, \apjl, 969, L22, \dodoi{10.3847/2041-8213/ad54c5}

\bibitem[{Schlecker {et~al.}(2025)Schlecker, Apai, Affholder, Ranjan, Ferri{\`e}re, Hardegree-Ullman, Lichtenberg, \& Mazevet}]{Schlecker2025}
Schlecker, M., Apai, D., Affholder, A., {et~al.} 2025, The Astrophysical Journal, 987, 24

\bibitem[{{Schlecker} {et~al.}(2024){Schlecker}, {Apai}, {Lichtenberg}, {Bergsten}, {Salvador}, \& {Hardegree-Ullman}}]{Schlecker2024}
{Schlecker}, M., {Apai}, D., {Lichtenberg}, T., {et~al.} 2024, \psj, 5, 3, \dodoi{10.3847/PSJ/acf57f}

\bibitem[{Shields {et~al.}(2013)Shields, Meadows, Bitz, Pierrehumbert, Joshi, \& Robinson}]{Shields2013}
Shields, A.~L., Meadows, V.~S., Bitz, C.~M., {et~al.} 2013, Astrobiology, 13, 715

\bibitem[{Stark {et~al.}(2024{\natexlab{a}})Stark, Latouf, Mandell, \& Young}]{Stark2024b}
Stark, C.~C., Latouf, N., Mandell, A.~M., \& Young, A. 2024{\natexlab{a}}, Journal of Astronomical Telescopes, Instruments, and Systems, 10, 014005

\bibitem[{Stark {et~al.}(2015)Stark, Roberge, Mandell, Clampin, Domagal-Goldman, McElwain, \& Stapelfeldt}]{stark2015}
Stark, C.~C., Roberge, A., Mandell, A., {et~al.} 2015, The Astrophysical Journal, 808, 149

\bibitem[{Stark {et~al.}(2014)Stark, Roberge, Mandell, \& Robinson}]{Stark2014}
Stark, C.~C., Roberge, A., Mandell, A., \& Robinson, T.~D. 2014, The Astrophysical Journal, 795, 122

\bibitem[{Stark {et~al.}(2019)Stark, Belikov, Bolcar, Cady, Crill, Ertel, Groff, Hildebrandt, Krist, Lisman, {et~al.}}]{stark2019}
Stark, C.~C., Belikov, R., Bolcar, M.~R., {et~al.} 2019, Journal of Astronomical Telescopes, Instruments, and Systems, 5, 024009

\bibitem[{Stark {et~al.}(2024{\natexlab{b}})Stark, Mennesson, Bryson, Ford, Robinson, Belikov, Bolcar, Feinberg, Guyon, Latouf, {et~al.}}]{Stark2024}
Stark, C.~C., Mennesson, B., Bryson, S., {et~al.} 2024{\natexlab{b}}, Journal of Astronomical Telescopes, Instruments, and Systems, 10, 034006

\bibitem[{Tayar {et~al.}(2022)Tayar, Claytor, Huber, \& van Saders}]{tayar2022}
Tayar, J., Claytor, Z.~R., Huber, D., \& van Saders, J. 2022, The Astrophysical Journal, 927, 31

\bibitem[{{The LUVOIR Team}(2019)}]{LUVOIR_final_report}
{The LUVOIR Team}. 2019, arXiv e-prints, arXiv:1912.06219.
\newblock \doarXiv{1912.06219}

\bibitem[{{Traub}(2003)}]{Traub2003}
{Traub}, W.~A. 2003, in ESA Special Publication, Vol. 539, Earths: DARWIN/TPF and the Search for Extrasolar Terrestrial Planets, ed. M.~{Fridlund}, T.~{Henning}, \& H.~{Lacoste}, 231--239

\bibitem[{{Traub} \& {Oppenheimer}(2010)}]{Traub2010}
{Traub}, W.~A., \& {Oppenheimer}, B.~R. 2010, in Exoplanets, ed. S.~{Seager}, 111--156

\bibitem[{Tuchow {et~al.}(2024)Tuchow, Stark, \& Mamajek}]{Tuchow2024HPIC}
Tuchow, N.~W., Stark, C.~C., \& Mamajek, E. 2024, The Astronomical Journal, 167, 139

\bibitem[{{Turbet}(2020)}]{Turbet2020}
{Turbet}, M. 2020, arXiv e-prints, arXiv:2005.06512, \dodoi{10.48550/arXiv.2005.06512}

\end{thebibliography}

\end{document}